\documentclass[english, apj]{emulateapj}
\usepackage[T1]{fontenc}
\usepackage[latin9]{inputenc}
\usepackage{array}
\usepackage{rotating}
\usepackage{units}
\usepackage{textcomp}
\usepackage{amsmath}
\usepackage{amsbsy}
\usepackage{amstext}
\usepackage{graphicx}
\usepackage{url}
\usepackage{babel}
\usepackage[backref,breaklinks,colorlinks,citecolor=blue]{hyperref}
\usepackage[all]{hypcap}

\makeatletter

\newcommand{\msun}{\mbox{$\,{\rm M}_\odot$}}

\begin{document}

\title{Exploding Satellites -- The Tidal Debris of the Ultra-Faint Dwarf Galaxy Hercules}

\author{Andreas H.W. K\"upper\altaffilmark{1,2}, Kathryn V. Johnston\altaffilmark{1}, Steffen Mieske\altaffilmark{3}, Michelle L.M. Collins\altaffilmark{4}, Erik J. Tollerud\altaffilmark{5}}
\altaffiltext{1}{Department of Astronomy, Columbia University, 550 West 120th Street, New York, NY 10027, USA}
\altaffiltext{2}{Hubble Fellow}
\altaffiltext{3}{European Southern Observatory, Alonso de Cordova 3107, Vitacura, Santiago, Chile}
\altaffiltext{4}{Department of Astronomy, Yale University, New Haven, CT 06511, USA}
\altaffiltext{5}{Space Telescope Science Institute, 3700 San Martin Dr, Baltimore, MD 21218, USA}
\email{Correspondence to: akuepper@astro.columbia.edu}

\begin{abstract}
The ultra-faint satellite galaxy Hercules has a strongly elongated and irregular morphology with detections of tidal features up to 1.3\,deg (3\,kpc) from its center. This suggests that Hercules may be dissolving under the Milky Way's gravitational influence, and hence could be a tidal stream in formation rather than a bound, dark-matter dominated satellite. Using Bayesian inference in combination with $N$-body simulations, we show that Hercules has to be on a very eccentric orbit ($\epsilon \approx 0.95$) within the Milky Way in this scenario. On such an orbit, Hercules ``explodes'' as a consequence of the last tidal shock at pericenter 0.5\,Gyr ago. It is currently decelerating towards apocenter of its orbit with a velocity of $V = 157$\,km\,s$^{-1}$ -- of which  99\% is directed radially outwards. Due to differential orbital precession caused by the non-spherical nature of the Galactic potential, its debris fans out nearly perpendicular to its orbit. This explains why Hercules has an elongated shape without showing a distance gradient along its main body: it is in fact a stream that is significantly broader than it is long. In other words, it is moving perpendicular to its apparent major axis. In this scenario, there is a spike in the radial velocity profile created by the dominant debris component that formed through the last pericenter passage. This is similar to kinematic substructure that is observed in the real Hercules. Modeling a satellite on such a highly eccentric orbit is strongly dependent on the form of the Galactic potential. We therefore propose that detailed kinematic investigation of Hercules and other exploding satellite candidates can yield strong constraints on the potential of the Milky Way.  
\end{abstract}


\section{Introduction}

Ultra-faint dwarf galaxies (UFDs) represent the extreme low-mass regime of galaxy formation. Their high inferred dynamical masses of $\gtrsim$10$^5$\msun\ to 10$^7$\msun, combined with their low luminosities ($-7\,\mbox{mag}\lesssim M_V\lesssim -1.5$\,mag) suggest that these objects are the most dark matter (DM) dominated systems in the local Universe \citep{Simon07, Geha09}. This makes them ideal targets for indirect DM searches, and extremely valuable for our understanding of structure formation in the Universe (e.g., \citealt{Tollerud08, Simon11, Collins13, Tollerud14}).

From their large half-light radii and their relatively low dynamical masses it can be inferred that the two-body relaxation times of most UFDs are several to hundreds of Hubble times. As such, their completely un-evolved dynamical states and their extremely low metallicities \citep{Kirby08, Ho15} make them ideal laboratories for studying star formation in low density and low metallicity environments \citep{Geha13}. 

However, deep imaging of some of these objects revealed that they may not be in dynamical equilibrium (see \citealt{Roderick16} and references therein), and instead may have experienced strong tidal interactions with their massive hosts, making our inferences of both their masses and stellar mass functions misleading (e.g., \citealt{Sand09, Penarrubia09, Sand12}). Here we are going to investigate the most elongated and most obviously disrupting UFD: Hercules (Herc). Discovered by \citet{Belokurov07} in SDSS data, Herc is a prime example of an object with an unclear dynamical state: deeper and wider observations reveal more and more signs of tidal disruption \citep{Coleman07, Sand09}. 

Recently, \citet{Roderick15} detected tidal features of Herc out to more than one degree distance from its center, corresponding to more than 2\,kpc at the heliocentric distance of Herc. These tidal features indicate that Herc may be significantly less massive than inferred from kinematics under the assumption of virial equilibrium. The measured mass-to-light ratio may therefore be inflated, and Herc may in fact have no dark matter component. The question is then: is Herc a galaxy at all, and if so, how can we tell? \citet{Brown14} analyzed deep HST imaging of Herc and other UFDs. They found that Herc's stellar populations resembles the old and metal-poor stars of the Milky Way globular cluster M\,92. However, while having an overall low metallicity comparable to M\,92, the observed spread in metallicities is higher than for most globular clusters (GCs), and only comparable to the very massive (and disputed) GCs Omega Centauri and M\,54 \citep{Willman12}. Based on these observations, Herc should be classified as a dwarf galaxy and not as a GC.

Dwarf galaxy or GC, both are subject to tidal forces and can be disrupted through energetic tidal shocks (e.g., \citealt{Johnston02, Penarrubia08, Klimentowski09}). Especially satellites falling into the gravitational potential of their hosts on nearly radial orbits will experience periodic, explosive mass loss on short timescales, but may yet be long-lived objects due to the orbital period being of the order of $>1$\,Gyr. Herc may be undergoing such explosive disruption. 

In fact, it has been suggested several times that Herc may be a disrupting or even completely unbound structure \citep{Zolotov11, Smith13} as it shows striking similarity to numerical models of disrupting satellites \citep{Kroupa97, Klessen98}. Most of these arguments were based on Herc's elongated and irregular shape \citep{Coleman07}. \citet{Simon07} found velocity substructure with a dispersion of about 1\,km\,s$^{-1}$ and a small offset of about 2\,km\,s$^{-1}$ from the assumed systemic velocity of Herc and suggested it could be due to disruption. However, such a velocity substructure was not a natural prediction of these dissolution scenarios. Moreover, scenarios in which Herc is interpreted as a tidal stream in formation predict a strong distance gradient across the body of Herc, for which there is no indication in the photometric data.

Here, we are going to demonstrate how the observational data can be explained within a scenario in which Herc is on a very radial orbit of eccentricity $\approx 0.95$. The elongated part of Herc that was found in SDSS and LBT data can then be interpreted as an \textit{exploded component} that has been blasted off of the Herc progenitor during the last pericenter passage. This component is being deflected asymmetrically by the gravitational potential of the Milky Way due to differential orbital precession. The particular geometry of this debris structure may yield powerful constraints on the shape of the Galactic gravitational potential.

We organized this investigation in the following way: In Sec.~\ref{sec:observations} we describe the available observational data and the constraints they put on models of Herc. Then in Sec.~\ref{sec:streakline} we derive an orbit for Herc using streakline modeling and Bayesian inference. In Sec.~\ref{sec:simulations} we study the behavior of satellites on such an extreme orbit by means of $N$-body simulations. The interpretation of this $N$-body data and the role of differential orbital precession for the appearance of Herc's particular features are given in Sec.~\ref{sec:differential}. Other signatures resulting from this scenario are described in Sec.~\ref{sec:signatures}. We conclude that Hercules is most probably a nearly-dissolved dwarf galaxy, which we see right now approaching its apocenter, and motivate future observations (Sec.~\ref{sec:conclusions}).

\section{Constraints on Hercules and its orbit}\label{sec:observations}
The Milky Way satellite Hercules was first discovered by \citet{Belokurov07} in data from the Sloan Digital Sky Survey. Its location on the sky is (RA, Dec$)=(247.77, 12.79)$\, deg, and (l, b$) = (28.73, 36.86)$\,deg in Galactic coordinates. Since its discovery, several follow-up observations and modeling attempts have tried to explain Herc's nature and current dynamical state.

\subsection{Observational constraints}

Various distance measurements put Herc far beyond the Galactic bulge into the very outskirts of the Milky Way halo. \citet{Belokurov07} determined a distance of $140^{+13}_{-12}$\,kpc from SDSS and INT data, while \citet{Sand09} derived a heliocentric distance of $133\pm6$\,kpc with LBT imaging data. \citet{Aden09b} used the apparent magnitude of the horizontal branch to estimate a distance of $147^{+8}_{-7}$\,kpc, and more recently, \citet{Brown14} used deep HST/ACS imaging of Herc's core to measure a distance of $(141\pm3)$\,kpc. The weighted mean of these estimates is 140\,kpc, which makes Herc one of the outermost (of the known) satellites of the MW.

Herc's total mass is largely unknown, which is due to the fact that its dynamical state is unclear. Using deep, wide-field LBT data, \citet{Coleman07} showed that Herc's surface density profile is very elongated, having an axis ratio of 3:1. Its ellipticity appears to vary from the innermost core having $e\approx0.3$ to the outermost parts of the satellite at about 10\,arcmin radius with $e\approx0.65$. \citet{Coleman07} already argued that this could be a sign of ongoing tidal disruption, but point out that Herc would have to be on a very eccentric orbit for that. With additional LBT imaging, \citet{Sand09} found a similar ellipticity of $e=0.67\pm0.03$, and measured a half-light radius of about 230\,pc. Moreover, they found over-densities of Herc-like stars up to about 1.3\,kpc from its center. These tidal feature are also traced by Blue Horizontal Branch stars \citep{Deason12}. The full extent of Herc's disruption has recently been demonstrated by \citet{Roderick15}: the authors used CTIO/DECam imaging data to reveal tidal debris of Herc out to a radius of more than 2\,kpc along the major axis of the satellite, and also perpendicular to this elongation (Tab.~\ref{tab:overdensities}). They find that there are at least as many Herc-like stars within its debris than in its main body, suggesting that Herc is losing mass at a high rate. 

\citet{Sand09} estimate the absolute integrated magnitude of Herc's main body to be $M_V = -6.2\pm0.4$\,mag, corresponding to a stellar mass of $\approx 10^4\msun$. Dynamical mass estimates, however, yield masses $>10^6\msun$. \citet{Simon07}, for example, use Keck/DEIMOS spectroscopy of essentially all targets within Herc that are bright enough to be observed from 10m-class telescopes to identify 30 likely Herc member stars. From these they measure a line-of-sight velocity dispersion of $\sigma=5.1\pm0.9$\,km\,s$^{-1}$. Assuming spherical symmetry, virial equilibrium, orbital isotropy, and making the assumption that Herc's mass follows its light distribution, the authors estimate a total mass of $(7.1\pm2.6)\times 10^6\msun$ and a dynamical mass-to-light ratio ($M/L$) of $332\pm221$. 

Contamination from foreground and background stars is a concern for the faint and distant satellite. \citet{Aden09b} use INT/WFC Str\"omgen photometry to clean their spectroscopic sample. Their final data set contains 18 red giant branch stars with radial velocity measurements (Tab.~\ref{tab:radialvelocities}). Those have a velocity dispersion of $\sigma=3.7\pm0.9$\,km\,s$^{-1}$, resulting in a dynamical mass of $3.7^{+2.2}_{-1.6}\times10^6\msun$ within a radius of 433\,pc, also assuming virial equilibrium, isotropy and spherical symmetry. Their estimated $M/L$ is with $103^{+83}_{-48}$ lower than the result of \citet{Simon07}, but still way beyond any $M/L$ of a simple stellar population.  Therefore, Hercules has to be either strongly dark-matter dominated, or brought out of virial equilibrium through tidal shocks.

The latter scenario is supported by the fact that despite its strong elongation, \citet{Simon07} find no sign of significant internal rotation in Hercules. \citet{Aden09b} do find a tentative radial velocity gradient, which may indicate some degree of rotation, but the gradient does not appear strong enough to suggest that Herc is a fully rotationally supported system. Moreover, its systemic velocity of $45.2\pm1.1$\,km\,s$^{-1}$ suggests that Hercules may be on a significantly radial orbit within the Milky Way \citep{Aden09b}: assuming that the Sun is in a right-handed, Galactocentric Cartesian coordinate system at $(-8.3, 0, 0)$\,kpc with a peculiar velocity of $(11.1, 254.3, 7.25)$\,km\,s$^{-1}$ \citep{Schoenrich12, Kupper15}, Herc's heliocentric radial velocity tells us that it is speeding away from the Milky Way with at least $140$\,km\,s$^{-1}$. Of course, this could either mean that Herc is on a very radial orbit or not bound to the MW at all. However, in an LCDM context, such un-bound satellites are extremely rare \citep{Boylan12, Tollerud14}

\subsection{Constraints from modeling Hercules}

\begin{table}
 \centering
 \label{tab:overdensities}
 \caption{Overdensities from Roderick et al. (2015)}
 \begin{tabular}{cccc}
 Segment & RA [deg] & DEC [deg] & Significance\\
 \hline
OD 6 & 247.51 & 12.34 & 3.74\\
OD 8 & 248.45 & 12.40 & 2.73\\
OD 9 & 248.15 & 12.51 & 2.33\\
OD 13.2 & 247.46 & 12.87 & 2.11\\
OD 13.3 & 247.37 & 12.98 & 0.74\\
OD 16.1 & 247.18 & 12.86 & 3.00\\
OD 16.2 & 247.09 & 12.92 & 1.29\\
OD 20 & 247.01 & 13.25 & 2.26\\
OD 23 & 247.12 & 13.47 & 3.15\\
OD 24 & 248.13 & 13.85 & 0.99\\
 \end{tabular}
\end{table}

\begin{table}
 \centering
 \label{tab:radialvelocities}
 \caption{Radial velocities from Ad\'en et al. (2009b)}
 \begin{tabular}{ccccc}
 ID & RA [deg] & DEC [deg] & $V_R$ [km\,s$^{-1}$]& $\Delta V_R$ [km\,s$^{-1}$]\\
 \hline
40222 & 247.93 & 12.78 & 41.90 & 3.65\\
40993 & 247.85 & 12.76 & 40.22 & 3.58\\
41082 & 247.85 & 12.75 & 41.73 & 0.67\\
41371 & 247.82 & 12.83 & 46.11 & 3.22\\
41423 & 247.81 & 12.79 & 34.79 & 5.06\\
41460 & 247.81 & 12.76 & 48.20 & 3.91\\
41642 & 247.79 & 12.70 & 48.09 & 8.30\\
41743 & 247.78 & 12.80 & 46.29 & 0.95\\
41758 & 247.78 & 12.80 & 43.18 & 3.51\\
41912 & 247.77 & 12.77 & 42.81 & 2.78\\
42096 & 247.75 & 12.82 & 54.61 & 1.62\\
42149 & 247.75 & 12.79 & 44.48 & 0.87\\
42170 & 247.74 & 12.76 & 32.72 & 4.77\\
42324 & 247.73 & 12.77 & 45.94 & 2.18\\
42637 & 247.70 & 12.82 & 49.41 & 3.16\\
42692 & 247.70 & 12.76 & 49.90 & 1.45\\
42795 & 247.68 & 12.83 & 42.90 & 1.57\\
43688 & 247.59 & 12.86 & 48.96 & 2.14\\
 \end{tabular}
\end{table}

Due to Herc's large heliocentric distance, its full spatial motion can only be inferred indirectly. \citet{Martin10} used the tentative radial velocity gradient across Herc's body of $16\pm3$\,km\,s$^{-1}$\,kpc$^{-1}$ found by \citet{Aden09a} to fit an orbit to the observational data, their main assumption being that Herc is not a self-gravitating dwarf galaxy anymore, but a tidal stream in formation. Hence, the orbit of the satellite, they argue, should be aligned with the major axis of the stellar distribution. With this simple model and in their assumed Galactic potential, the authors found a tangential velocity of only $V_T = -16^{+6}_{-22}$\,km\,s$^{-1}$ in the Galactic rest frame. Such a low velocity puts Herc on a very radial orbit with an eccentricity of $\epsilon = 0.95$ and with a pericenter of $R_{P} = 6^{+9}_{-2}$\,kpc -- in agreement with the expectations for a tidally shocked, dissolving satellite.

\citet{Blana15} use this suggested orbit for Hercules for an efficient, systematic search of an $N$-body model that reproduces Herc's present-day faint and extended appearance. For this purpose, the authors create mock observations of their simulations, measuring the mass, central surface brightness, projected half-light radius, velocity dispersion and velocity gradient of each model. Within their chosen Galactic potential they succeed in finding models that reproduce these basic features of Herc. However, their models fail to reproduce its ellipticity and orientation of the major axis on the sky simultaneously -- \citet{Blana15} observe that, when the ellipticity of the remnant reaches the observed value of $\approx0.67$, the orientation ``flips'' by about 100\,deg and misaligns with Herc's orbit.

\citet{Blana15} try to understand this interesting behavior. They identify three regimes for the state of the model at the end of the simulations: 
\begin{enumerate}
\item a ``bound regime'' in which the central, bound object appears nearly spherical and largely unaffected by tides. In the bound regime, the object is surrounded by low-density debris,\\
\item a ``tidal regime'' in which the satellite has lost large parts of its mass, and appears elongated along its orbit. The satellite remnant is surrounded by lots of debris in this stage,\\
\item a ``stream regime'' in which the satellite is completely disrupted and the debris is spread along the progenitor's orbit.\\
\end{enumerate}
\citet{Blana15} observe the ``flip'' in orientation of the debris when the satellite has been completely destroyed in the last pericenter passage, that is, when it is in between the ``tidal'' and the ``stream'' regime. Interestingly, their best-fit models for the other observational constraints are also in this intermediate regime, where Herc is right on the edge of destruction. 

Since none of their models reproduces all of Hercules' observational constraints, \citet{Blana15} argue that the orbit they chose is most likely wrong. This conclusion appears reasonable, given that \citet{Martin10} assumed that the orientation of the elongation of the debris must follow the orbit of the progenitor. As we will show in the following section, a similar orbit for Hercules can be found if this assumption is dropped (Sec.~\ref{sec:streakline}). Using $N$-body simulations, we will show that Herc's debris spreads nearly perpendicular to its orbit if the satellite is in between the tidal and the stream regime at the present day (Sec.~\ref{sec:simulations}). In Sec.~\ref{sec:differential} we will then explain that this spread is due to differential orbital precession of debris stars and outline how this effect can be used to infer the non-spherical nature of the Galactic potential.

\section{Hercules' orbit from streakline modeling}\label{sec:streakline}

Long tidal streams often align very well with the orbit of their progenitors. However, this is not the case in general, and tidal streams are not always elongated structures. In fact, tidal streams can be significantly misaligned from the orbits of their progenitors \citep{Klimentowski09, Sanders13}, and the alignment of stream elongation and orbit can even be lost completely. This is especially then of importance when the progenitor is on an eccentric orbit and approaching its apocenter \citep{Kupper10}. The compression of the debris into the vicinity of the progenitor is highest in this orbital phase, where stream and progenitor can even become indistinguishable \citep{Kupper11a, Hendel15}. In such cases, the stream-progenitor system can appear as a ``cloud'' of stars, or the dense center of the satellite may be regarded as having a debris ``halo'' \citep{Olszewski09}. 
 
Since we are not sure what the full extent of Herc's debris is, 
it appears reasonable to not impose any prior on the elongation angle of the debris with respect to the orbit. In order to determine its orbit without using this constraint, we use the \textit{streakline} methodology presented in \citet{Kupper15}, where it was applied to the debris of the Milky Way globular cluster Palomar\,5. This cluster shows a long, thin tidal stream in SDSS data, which seems to closely follow the cluster's orbit. However, the streakline method makes no explicit assumption on this alignment and can therefore deal with any form of debris.

\subsection{Streakline model setup}\label{sec:streaklinesetup}

Streakline modeling uses restricted three-body integrations of test particles to simulate the shape of a satellite's debris \citep{Varghese11, Kupper12, Bonaca14}. Starting from the present-day position of the satellite, the orbit is integrated backwards in a trial galactic potential for a given amount of time. Then, from this past position, the orbit is integrated back to the present-day while test particles are released from the instantaneous Lagrange points (L1 and L2) of the satellite-galaxy system at fixed time intervals. The satellite and the test particles are then integrated to the present-day and their on-sky projections are compared to the shape of the satellite's observed debris.

For this purpose the orbit of the satellite has to be either known or inferred from matching the streakline models to the observed parts of the debris. In the latter case, the orbit is drawn from a set of parameters including its six present-day phase space coordinates. But, generally, the shape of the debris also depends on the progenitor's mass, its mass-loss rate and, most importantly, on the shape of the Galactic potential.

For some nearby satellites like the Milky Way globular cluster 47\,Tucanae, the full 6D phase-space information of the progenitor is available from observations and we can predict the shape of its debris simply by using these best estimates in the integrations \citep{Lane12}. In most cases, however, we have as little information as for Herc: some components are well determined, whereas others are missing completely. 

Given the available data, we use the present-day position, heliocentric distance (140\,kpc), and radial velocity (45.2\,km\,s$^{-1}$) of Herc from the literature as described above. We leave the two missing proper motion components of Herc's velocity as free parameters and use a Bayesian approach in combination with Markov chain Monte Carlo inference to find their most likely values (see Sec.~\ref{sec:bayesian}). We also leave Herc's present-day mass and its mass-loss rate as free parameters. We allow all model parameters to vary widely (see Tab.~\ref{tab:priors}) by using flat priors that were given through the observational constraints summarized in Sec.~\ref{sec:observations}.

Since we are interested in finding a model of Herc in which it is disrupted by strong tidal forces, we assume that there is significant random scatter in the escape conditions of stars from the main body. As with the modeling of Palomar\,5 in \citet{Kupper15}, we assume that stars are preferentially stripped from the Lagrange points of the progenitor. However, in contrast to Palomar\,5, the escape conditions for stars from Herc are given random spatial and velocity offsets around these kinematically cold escape conditions. This is based on the assumption that mass loss is driven by tidal stripping rather than two-body relaxation. This process is more violent and results in more scatter in the escape conditions of stars from the progenitor, and hence in kinematically hotter tidal tails. Therefore, we pick a strong spatial scatter following a Gaussian distribution with a dispersion of 0.25 times the size of the instantaneous tidal radius from the locations of the two instantaneous Lagrange points L1 and L2. The velocity offsets from the progenitor are chosen such that they match the progenitor's angular velocity with a random Gaussian velocity offset with a dispersion of 2\,km\,s$^{-1}$. These escape conditions produce tidal stream models that are similarly spread out like the ``kinematically hot'' stream models of 47\,Tuc's tidal tails in \citet{Lane12}. Ultimately, we found that the detailed choice of this scatter is not important when we are only interested in the overall shape of the debris, since the spread in the debris is largely dominated by the effects of differential orbital precession.

We do not have any information about the age of the debris. But generally, we can say that the older the debris is, the wider spread out it will be. For this reason, we have to leave the integration time, $t_{int}$, of the streakline models as a free parameter. As a consequence of this, the number of streakline model particles will vary from model to model as we keep the time interval fixed to 1 test particle per Myr. In order to avoid biases towards shorter or longer integrations times, we will normalize our likelihoods accordingly (see Sec.~\ref{sec:bayesian}).

The gravitational potential of the Milky Way is not well determined out to the galactocentric radius of Hercules. Thus, we should fit the debris by varying Herc's orbital parameters, its mass and mass-loss rate, as well as the shape of the potential of the Galaxy. However, the observational constraints on Herc are not sufficient for such a complex approach. The same holds true for a more complex model of Herc with additional model parameters for, e.g., rotation of the progenitor \citep{Amorisco15} or for an orbital phase-dependent mass loss rate \citep{Fardal15}. Given the low number of observational constraints, we would likely be over-fitting our data without getting meaningful constraints on the model parameters. For this reason, we keep the model of Herc simple and fix the potential of the Milky Way to the best-fit values from \citet{Kupper15}, which are described in Sec.~\ref{sec:setup}. With this in mind, we will restrict ourselves to qualitative arguments here and restrain from making quantitative statements on Herc's exact orbit or internal state.

A summary of the five model parameters and the ranges of the flat priors that we assume can be found in Table~\ref{tab:priors}.

\subsection{Bayesian modeling framework}\label{sec:bayesian}
\begin{table}
 \centering
 \label{tab:priors}
 \caption{Summary of model parameters}
 \begin{tabular}{lll}
 Parameter & Range of flat prior & Median and posterior width\\
 \hline
$M_{Herc}$ & $[0, 2\times10^5]\msun$ & $(5.1^{+9.8}_{-4.1})\times10^4\msun$ \\

$dM/dt$ & $[0, 200] \msun\,$Myr$^{-1}$ & $72.2^{+92.6}_{-50.0} \msun\,$Myr$^{-1}$ \\

$\mu_{\alpha} \cos(\delta)$ & $[-1, 1]$\,mas\,yr$^{-1}$ & $-0.210^{+0.019}_{-0.013}$\,mas\,yr$^{-1}$ \\
 & $[-668, 668]$\,km\,s$^{-1}$ & $-139^{+13}_{-9}$\,mas\,yr$^{-1}$ \\
$\mu_\delta$ & $[-1, 1]$\,mas\,yr$^{-1}$ & $-0.224^{+0.015}_{-0.016}$\,mas\,yr$^{-1}$ \\
 & $[-668, 668]$\,km\,s$^{-1}$ & $-149^{+10}_{-11}$\,mas\,yr$^{-1}$ \\
$t_{int}$ & $[-10,-1]$\,Gyr & $-3.2^{+1.2}_{-1.7}$\,Gyr
 \end{tabular}
\end{table}

We generate streakline models of Hercules's tidal debris and compare their distribution to the observed overdensities of Herc's debris and to the locations and velocities of stars surrounding Herc. To assess the quality of the fit of the model to the data and to guide our Markov Chain Monte Carlo sampler, we use a Bayesian approach to calculate the likelihood of the data based on the individual models. Following \citet{Bonaca14}, we define a likelihood function, $P$, that describes the probability of the $N$ observational data points, $X_n$, given the $K$ model particles, 
\begin{equation}
P\left(\{X_n\} | \theta,I\right)=\prod_{n=1}^{N}p\left(X_n| \theta, I\right),
\end{equation}
where $\theta$ is a set of model parameters and $I$ are the priors on these parameters. Each data point, overdensity or star with measured radial velocity, has an independent likelihood of being produced by the model. We can therefore combine all likelihoods in a product. The individual likelihoods of each overdensity or radial velocity star being produced by the model points is given by
\begin{equation}
p\left(X_n | \theta,I\right)=\sum_{k=1}^{K}P_kp\left(X_n| k,\theta, I\right),
\end{equation}
where we marginalized over all $K$ model points, with the normalization factor $P_k = 1/K$. The individual probabilities of each data point, $X_n$, being produced by a given model particle, $x_k$, can be estimated using the measurement uncertainties of the data points, $\sigma_n$, and assuming that the probability distribution of each overdensity or radial velocity star is given by a Gaussian distribution, $\mathcal{N}$, centered on the model particle, $x_n$. The likelihood can then be written as 
\begin{equation}
p\left(X_n | x_k,\theta,I\right)=\mathcal{N}\left(X_n| x_k,\sigma_n^2 + \Sigma^2_k\right).
\end{equation}
Here, $\Sigma_k$ is a smoothing tensor, which we use in order to construct a smooth distribution function out of the discrete model points. They are initially set to small random values, and kept as free hyper-parameters with wide, flat priors during inference. For a more detailed description of this framework, see \citet{Kupper15}.

For the comparison of the streakline models to observational data, we use the positions of the debris clumps of Herc as detected by \citet{Roderick15} that are listed in Tab.~\ref{tab:overdensities}. Furthermore, we add the cleaned set of radial velocities from \citet{Aden09b} as observational constraints (Tab.~\ref{tab:radialvelocities}). With this likelihood function and the streakline model parameters described in Sec.~\ref{sec:streaklinesetup}, we use the publicly-available Markov Chain Monte Carlo sampler \texttt{emcee} \citep{Foreman13} to perform inference. 

Similar to \citet{Kupper15}, we use \texttt{emcee} in parallel-tempering mode with two temperatures, and a setup of 128 walkers in each of temperature. After a burn-in phase, we let each walker take 800 steps, resulting in a sample of $10^5$ likelihood evaluations. We then visually inspected the chains for convergence. We present the posterior probability distributions for the model parameters in the following section.

\subsection{Modeling results for Hercules and its orbit}\label{sec:orbit}

\begin{figure*}
\centering
\includegraphics[width=0.45\textwidth]{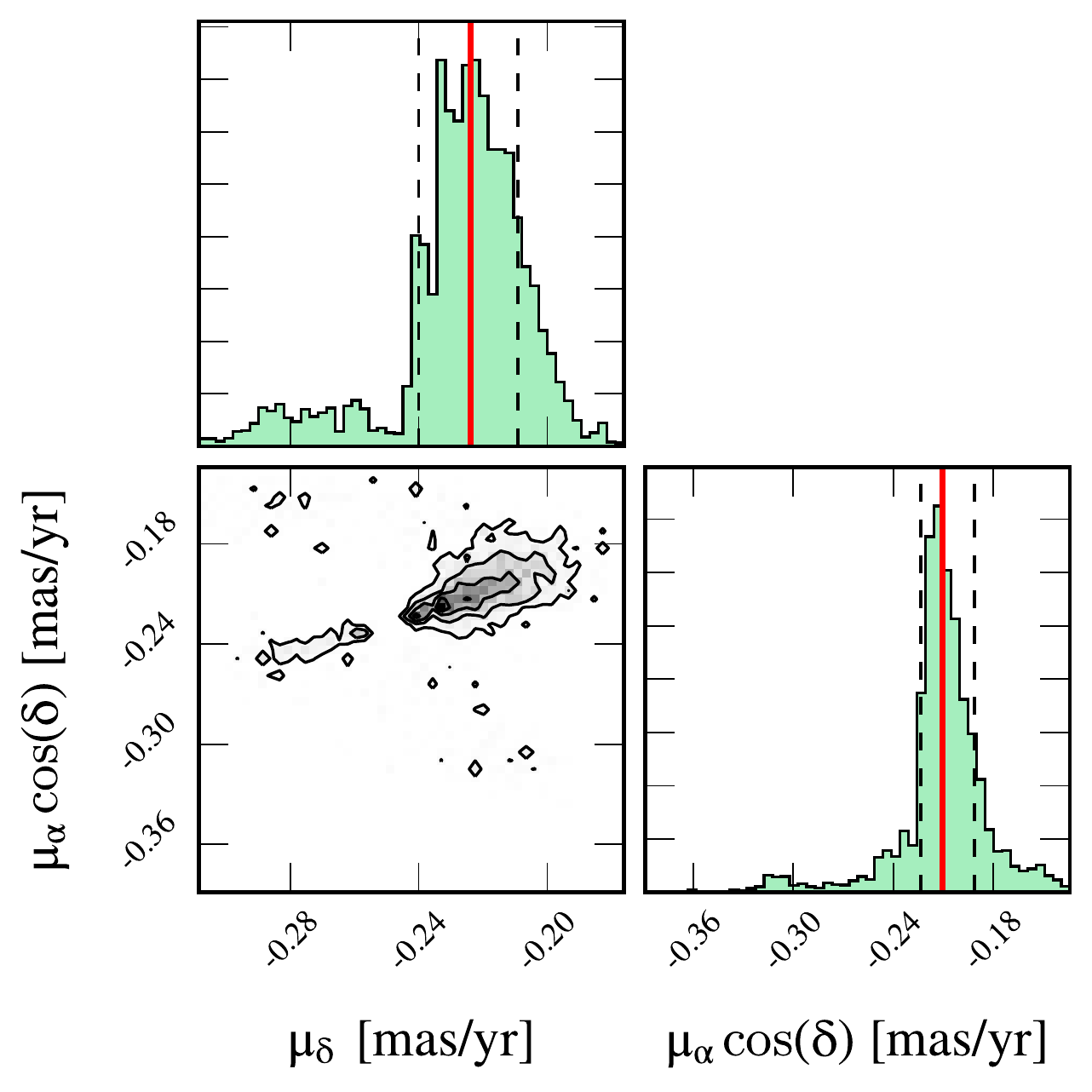}
\includegraphics[width=0.45\textwidth]{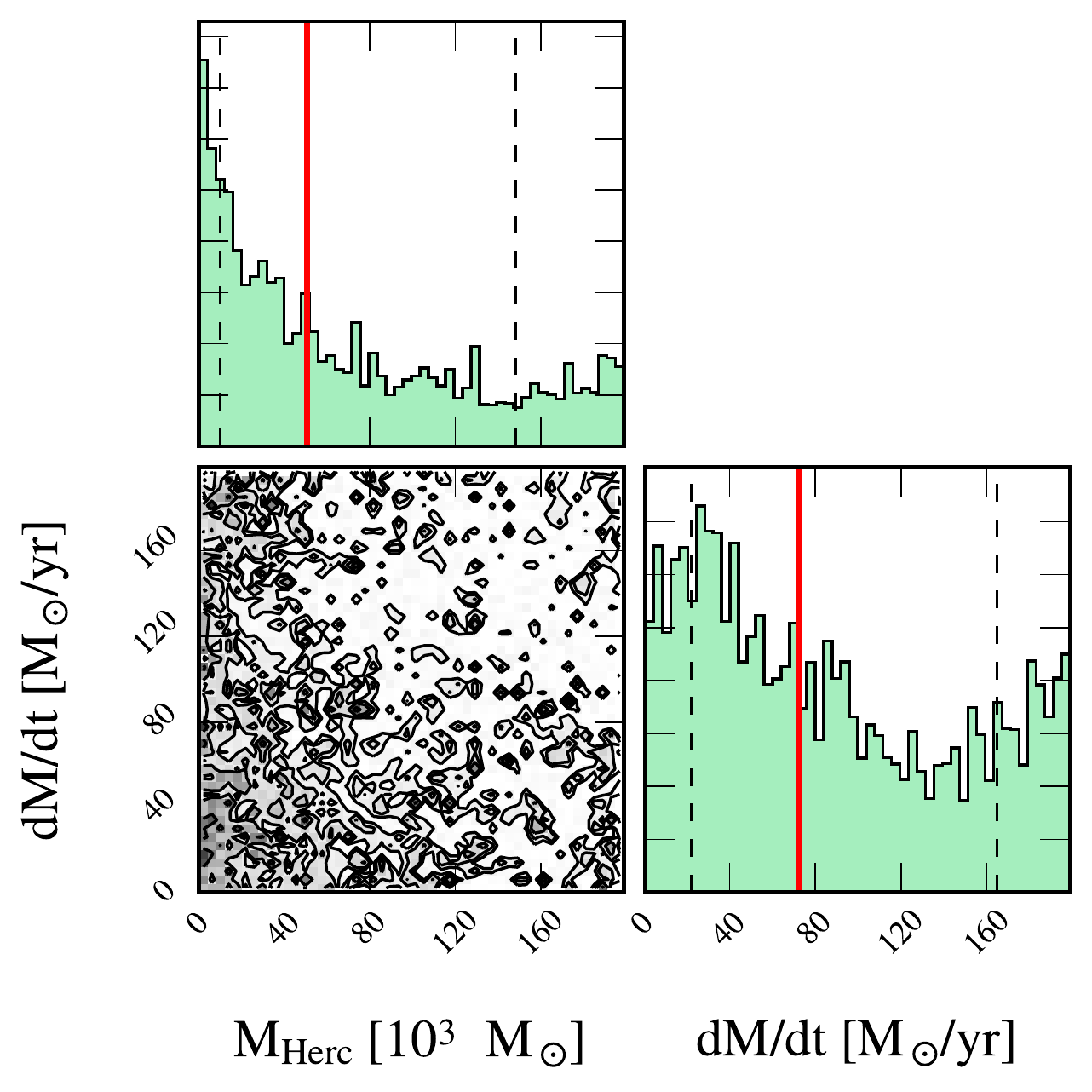}
\caption{Posterior distributions of four out of five Hercules-model parameters that were allowed to vary freely in the MCMC run (integration time is not shown). \textbf{Left:} due to Herc's large heliocentric distance and peculiar shape, the proper motion components are strongly constrained by the observational data to be $\mu_\alpha\cos(\delta)=(-0.21\pm0.01)$\,mas\,yr$^{-1}$ and $\mu_\delta=-0.22^{+0.01}_{-0.02}$\,mas\,yr$^{-1}$. \textbf{Right:} Hercules' present-day mass, $M_{Herc}$, is largely unconstrained, but shows a tendency towards masses around zero. Its time-averaged mass-loss rate $dM/dt$ is not constrained by the data, and hence shows a basically flat posterior distribution.}
\label{mcmc}
\end{figure*}

\begin{figure*}
\centering
\includegraphics[width=0.45\textwidth]{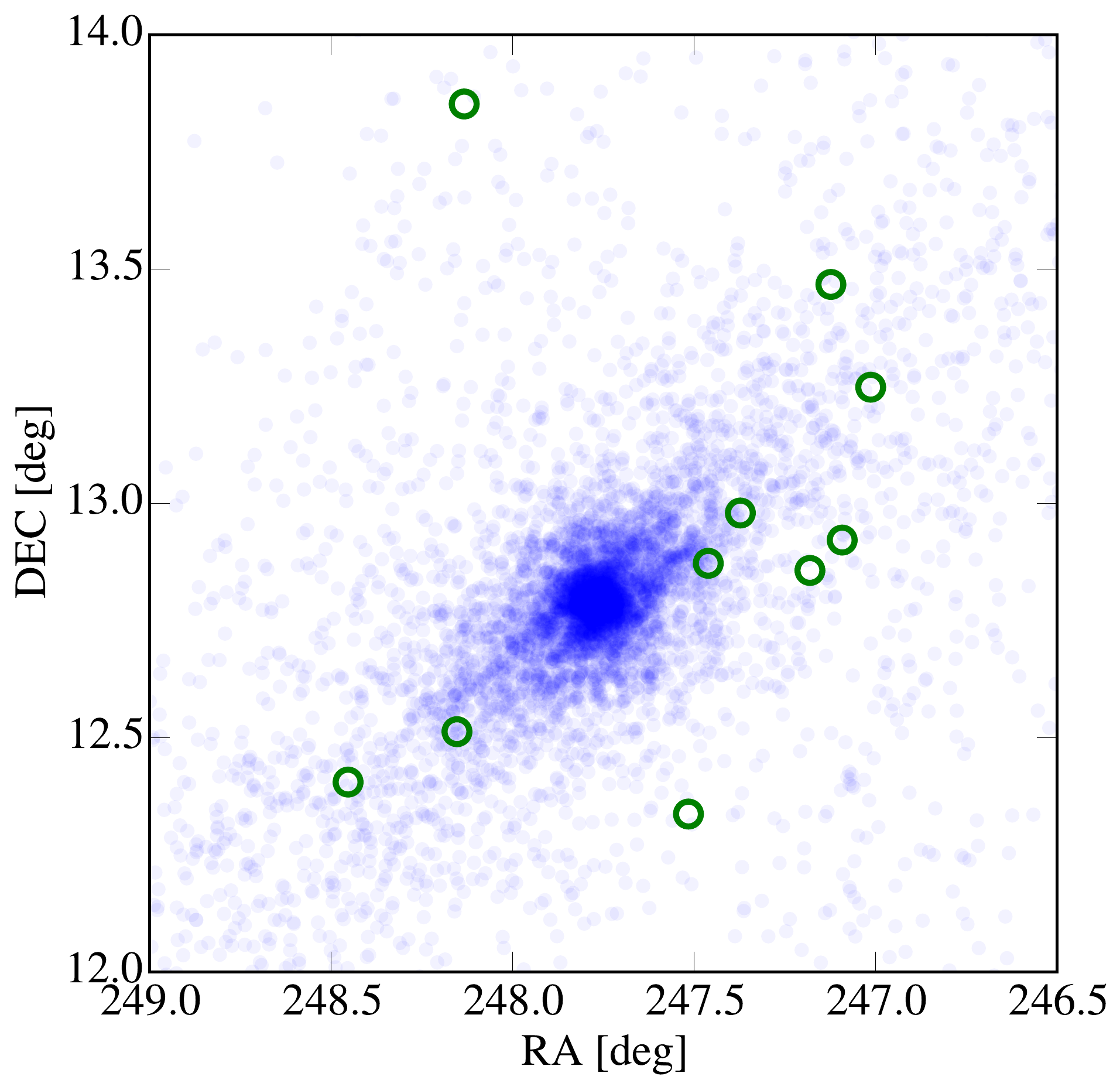}
\includegraphics[width=0.44\textwidth]{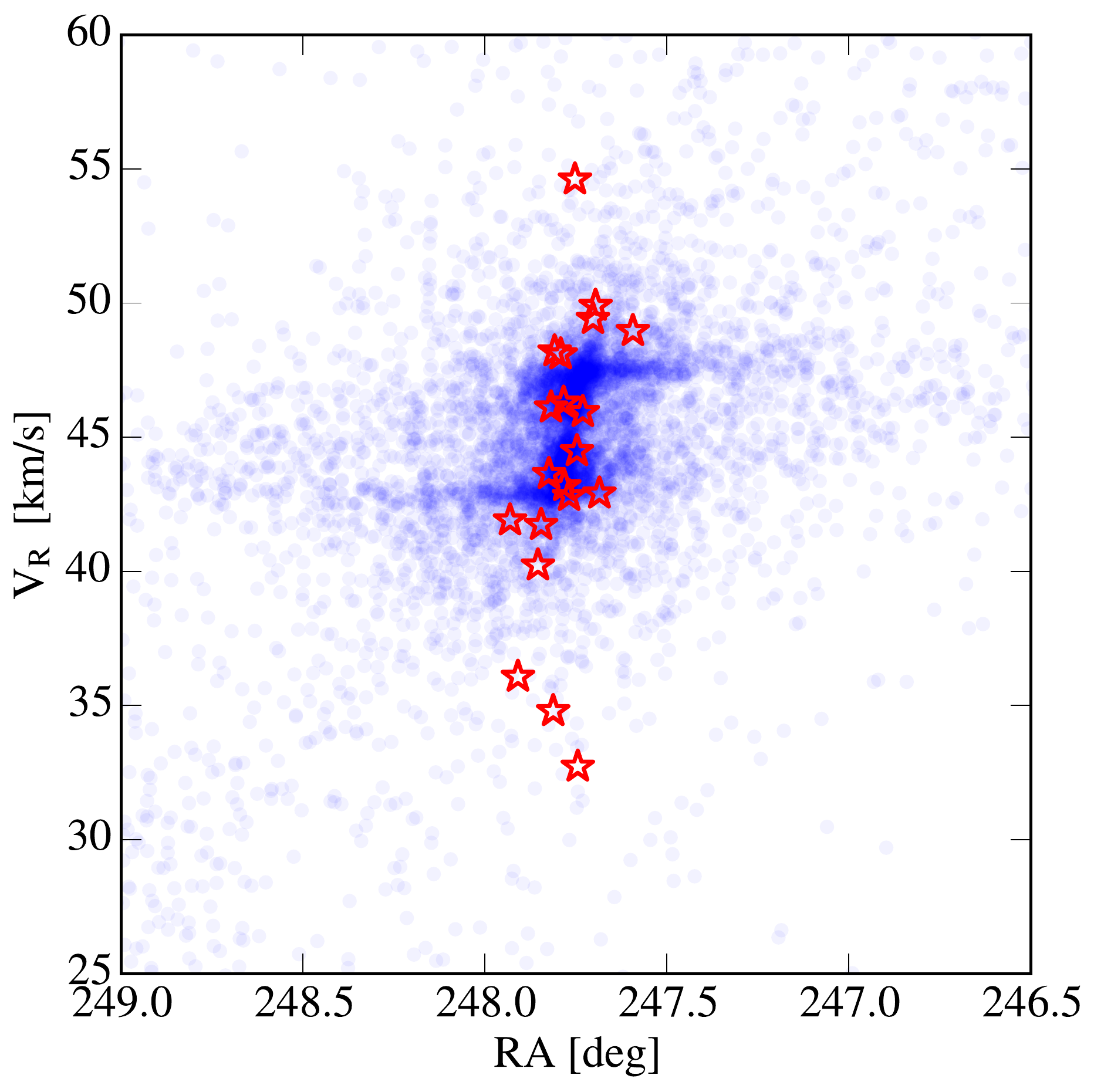}
\caption{Best-fit model from the MCMC run using the median values from the posterior parameter distributions. Blue, semi-transparent points show the streakline model, whereas green circles in the left-hand panel mark the positions of tidal debris as detected by \citet{Roderick15}. Red star symbols in the right-hand panel show positions of radial velocity measurements from \citet{Aden09a}. The simple streakline model reproduces the elongated debris structure and the tentative radial velocity gradient, but also shows that Herc may be surrounded by lots of diffuse tidal debris.}
\label{medianmodel}
\end{figure*}

\begin{figure*}
\centering
\includegraphics[width=0.45\textwidth]{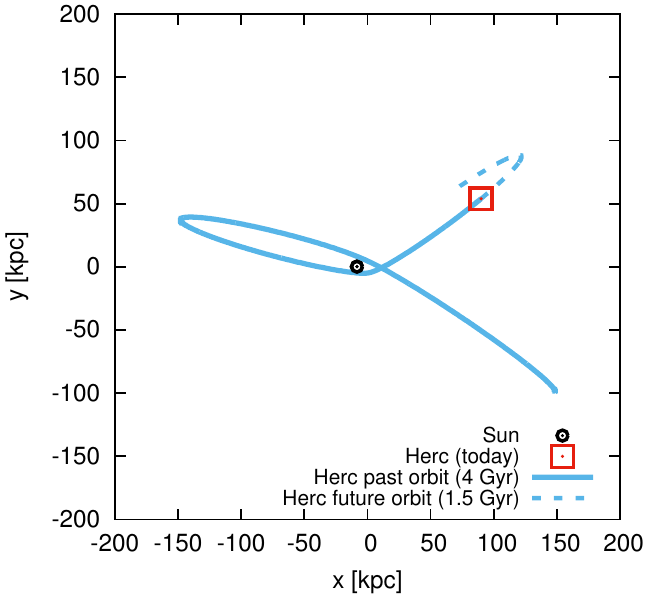}
\includegraphics[width=0.45\textwidth]{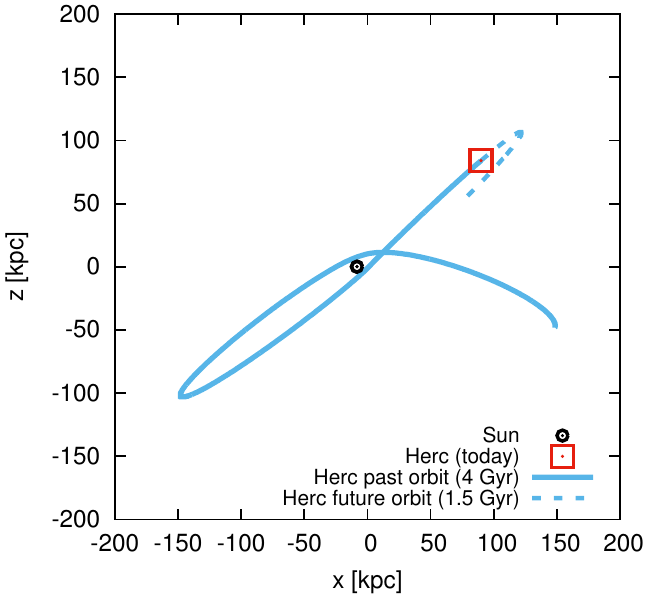}
\caption{Best-fit orbit resulting from the MCMC modeling predicts Hercules to be on a highly eccentric orbit ($\epsilon\approx0.95$) in the Galactocentric rest frame. The solid line shows the past 4\,Gyr of its orbit (left panel: projected onto the Galactic-disk plane; right panel: orbit perpendicular to the disk plane), and the dashed line shows the next 1.5\,Gyr. The last pericenter shock occurred about 0.5\,Gyr ago. Due to the eccentricity of the orbit, the time in and around apocenter is much larger than the rest of the orbit, making it very likely to observe the satellite and its debris in a tidally compressed state.}
\label{orbit}
\end{figure*}

Figure \ref{mcmc} shows the posterior probability distributions of four out of five model parameters. The proper motion components are strongly constrained, whereas Herc's present-day mass and mass-loss rate are completely unconstrained. Experimenting with the streakline modeling setup by, e.g., changing the degree of scatter in escape conditions or allowing more or less parameters to vary, produces similar results. We restrict ourselves to discussing only this one specific choice of model setup since we are mainly interested in the underlying effect that causes the Hercules debris to appear elongated and misaligned with the satellite orbit. Keeping these simplifications in mind, we find that, if Herc's elongation is caused by tidal disruption, it has to be on a very specific/fine-tuned orbit.

The type of orbit preferred by the data results in models like the one shown in Fig.~\ref{medianmodel}. The debris spreads out mostly along the SE/NW axis, but debris can be found in any direction of the main body. The models show a spread in radial velocities across the central part of the debris with a slight radial velocity gradient in right ascension as seen in the observational data.   

As summarized in Tab.\ref{tab:priors}, the median values of the proper motion posteriors, and their 16th and 84th percentiles are 
\begin{eqnarray}
\mu_{\alpha} \cos(\delta) &=& -0.210^{+0.019}_{-0.013}\,\mbox{mas}\,\mbox{yr}^{-1},\\
\mu_\delta &=& -0.224^{+0.015}_{-0.016}\,\mbox{mas}\,\mbox{yr}^{-1},
\end{eqnarray}
corresponding to tangential velocities of $139^{+13}_{-9}$\,km\,s$^{-1}$ and $149^{+10}_{-11}$\,km\,s$^{-1}$, respectively. For comparison, the allowed prior range corresponds to $\pm668$\,km\,s$^{-1}$. Combined with the (fixed) heliocentric radial velocity of 45.2\,km\,s$^{-1}$, and put into the Galactic rest frame, gives Herc a 3D space velocity of only about $V=157$\,km\,s$^{-1}$. Most of this velocity (155\,km\,s$^{-1}$) is directed radially away from the Galactic center, making Herc's orbit very radial. The remaining tangential velocity of about $V_T = 25$\,km\,s$^{-1}$ is similar to the 16\,km\,s$^{-1}$ that \citet{Martin10} got through their approach. However, the orientation on the sky of our orbit compared to theirs is nearly perpendicular (see Sec.~\ref{sec:structure}).

Fig.~\ref{orbit} shows the median orbit we get for Herc. The eccentricity of this orbit is $\epsilon=0.95$ with an apocenter of $R_A = 185$\,kpc and a pericenter of only $R_P = 4.9$\,kpc. We plot the orbit for the past 4\,Gyr, in which Herc has had two pericenter passages. The dashed line shows the next 1.5\,Gyr. Herc is currently approaching the apocenter of its orbit, but will not reach it within the next $\approx675$\,Myr. One orbit takes about 2.3\,Gyr, so about half of time Herc spends in an orbital phase as close to apocenter as it is now or even closer. From this point of view, it is statistically plausible to find disrupted objects like Herc in the outer halo of the Milky Way. 

The posterior probability distributions of the other parameters are less constraining. For the integration time, we get a median value of $t_{int} = 3.2$\,Gyr with an uncertainty of 1--2\,Gyr. This implies that most of the debris can be explained by material that got stripped from Herc's main body during the last one or two pericenter passages. 

As mentioned above, Hercules' mass and mass-loss rate remain basically unconstrained through our modeling. Given the large eccentricity of the orbit, the tidal forces of the Milky Way vary significantly along the orbit. The distance of the Lagrange points from the center of Herc, which is important for the initial spatial offsets of the test particles from the main body, is only weakly dependent on these two quantities. Hence, a large range of masses and mass-loss rates can create debris that matches the shape and extent of the Hercules observations. Important to note, though, is the fact that even models with a present-day mass of zero can reproduce the detected features of Hercules.

In order to better understand why this is, we use our best-fit orbit to run detailed $N$-body simulations of disrupting satellites, which we present in the following section.

\section{Hercules $N$-body simulations}\label{sec:simulations}

Our streakline modeling approach is very simplified and may not be realistic enough to model the disruption of a satellite on such an eccentric orbit. We therefore perform follow-up $N$-body simulations of the median orbit described in the previous section.

\subsection{Setup of the N-body simulations}\label{sec:setup}

We run 6 simulations of identical models that only differ in their radial extent. That is, we keep the satellite mass fixed but vary its initial half-mass radius. In this way we can test different degrees of susceptibility to tidal disruption. Models with a larger initial radius have a lower average density and are therefore easier disrupted during pericenter passages when the satellite passes through the high-density part of the Galaxy.

Our streakline modeling has shown that the currently known tidal debris features do not require the presence of a massive galaxy but can be explained by a disrupting progenitor with a present-day bound mass of less than $10^5\msun$ and likely close to zero (Fig.~\ref{mcmc}). We have also shown that the dispersion of debris is slow. That is, the whole structure of Herc and its debris that we see today can be explained by dispersed debris from about $3-4$\,Gyr of dynamical evolution. Therefore, our test models have an initial total mass of only 50\,000\msun, in agreement with the model choices by \citet{Blana15}. As we will show, this mass is sufficient to explain all features of the present-day Hercules. However, the satellite may have been significantly more massive at birth and may have been embedded in a massive dark matter halo.

The models consist of 50\,000 particles of 1\msun~each, which we set up as Plummer spheres using the publicly available code \texttt{McLuster}\footnote{\url{https://github.com/ahwkuepper/mcluster}} \citep{Kupper11b}. For simplicity, we assume that there is only one mass component. The composition of this mass is not specified, and could be either mass in stars or dark matter. To match the observed structure and brightness of Hercules, the final integrated surface brightness of simulation particles can be scaled to the mass-to-light ratio of the observed surface brightness (see, e.g., \citealt{Blana15}). We vary its initial half-mass radius from $40-90$\,pc in steps of 10\,pc, resulting in initial half-mass densities between about 0.2$\msun\,$pc$^{-3}$ and 0.02$\msun\,$pc$^{-3}$.    

For the numerical integration we use \texttt{NBODY6}\footnote{\url{https://github.com/nbodyx/Nbody6}} \citep{Aarseth03}, which has been adapted to GPU graphic cards using CUDA for high performance \citep{Nitadori12}. We set up the orbit of the satellite according to our best-fit results described in Sec.~\ref{sec:orbit}. The code has a three-component galaxy potential implemented, which consists of a spheroidal bulge component \citep{Hernquist90}, a \citet{Miyamoto75} disk component, and a dark-matter halo component. We modified the latter such that we can use an oblate NFW halo \citep{Navarro97} with a flattening in the potential perpendicular to the disk plane. In this way we can use the exact same galactic potential as for the streakline modeling.

Since the observational evidence is not sufficient to constrain the potential of the Milky Way, we decided to use the Galactic gravitational potential from \citet{Kupper15} as an exemplary test case. As bulge mass we use $3.4\times10^{10}\msun$, and a scale radius of 0.7\,pc. The disk component has a scale mass of $10^{11}\msun$, a disk scale length of 6.5\,kpc, and a scale height of 260\,pc. The halo component has a scale mass of $1.58\times10^{12}\msun$, a scale radius of 37.9\,kpc and a potential flattening of 0.95 (where 1 would be spherical). The choices for the parameters are the same as the ones we used for the streakline modeling. They are described in more detail in \citet{Kupper15}.   

The posterior probability distribution for the integration time of our streakline models indicates that the tidal debris, which was observed around Herc 0.25--1\,deg from the main body, was produced during the last one or two orbits. We therefore run the $N$-body simulations for 4\,Gyr, which means that the satellite starts off in apocenter of its orbit and goes through two pericenter passages before ending in its present-day position. All particles are kept in the simulations and are fully integrated throughout the runs.

\subsection{Evolution of the N-body models}

\begin{figure*}
\centering
\includegraphics[width=0.45\textwidth]{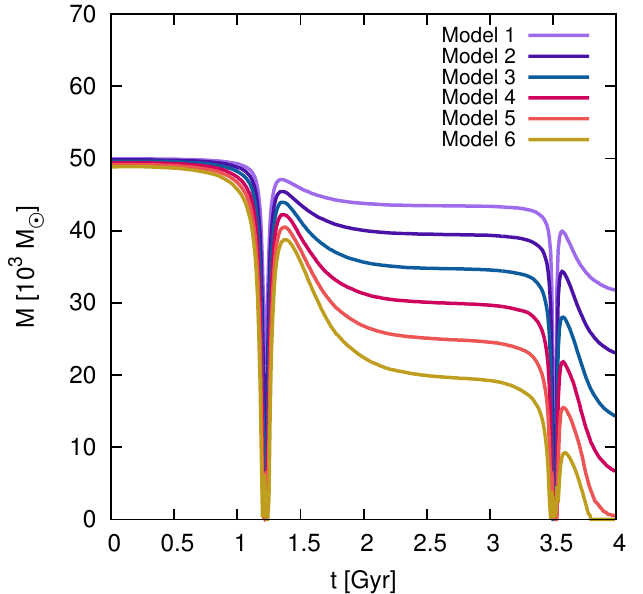}
\includegraphics[width=0.45\textwidth]{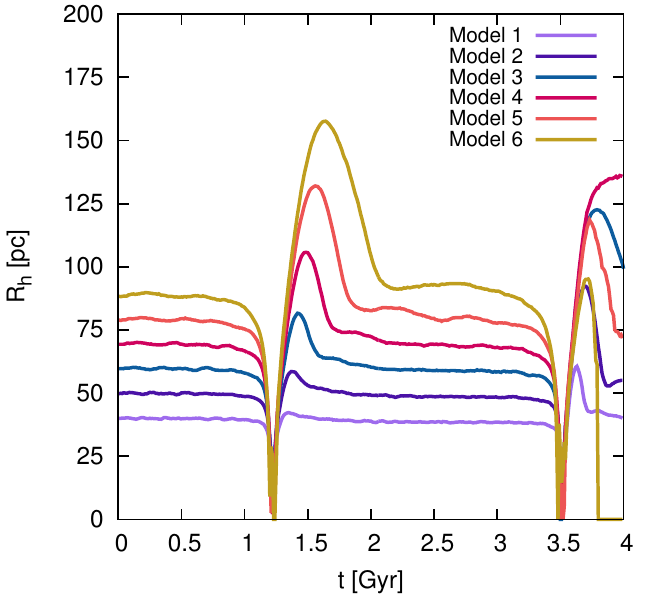}
\caption{Time evolution of the six $N$-body simulations of extended satellites on the orbit derived from streakline modeling in Sec.~\ref{sec:streakline}. The models have the same initial mass of $M_0 = 50\,000\msun$, but differ in their initial half-mass radii -- from 40\,pc (Model\,1) to 90\,pc (Model\,6). Left panel: bound mass, i.e.~mass within the Jacobi radius. Right panel: half-mass radius of the bound mass. The two pericenter shocks at $t=1.2$\,Gyr and $t=3.5$\,Gyr have profound impact on the satellites, leaving all stars momentarily unbound. Model\,6 gets completely disrupted by the final shock.}
\label{mass}
\end{figure*}

\begin{figure*}
\centering
\includegraphics[width=0.32\textwidth]{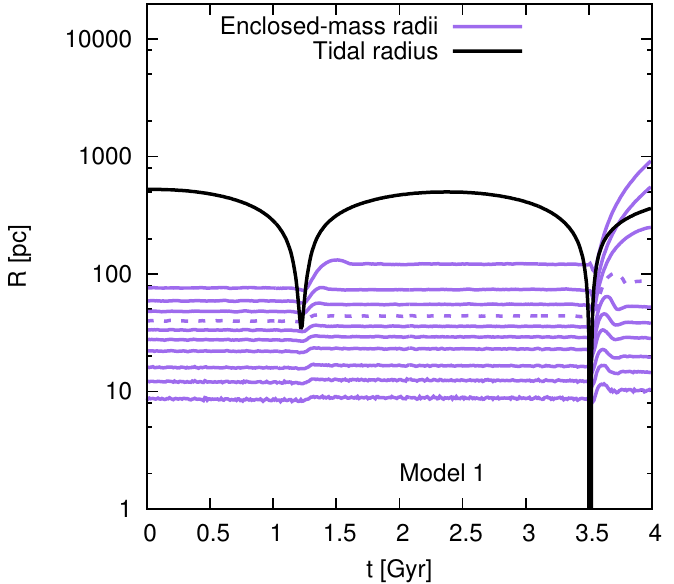}
\includegraphics[width=0.32\textwidth]{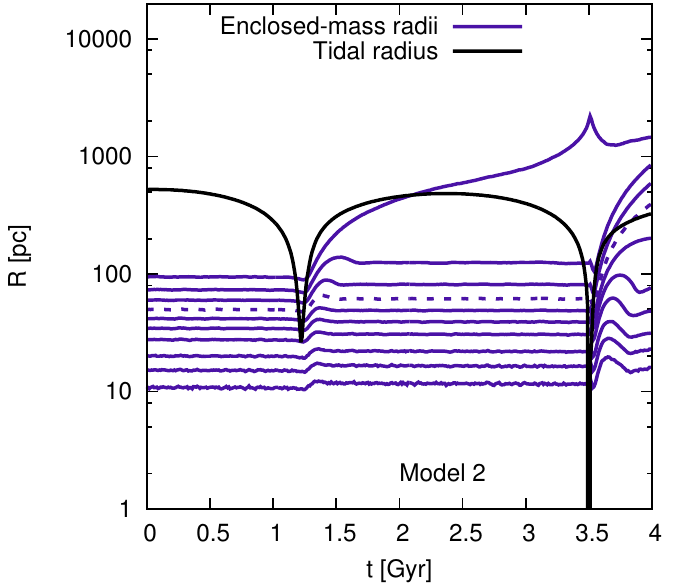}
\includegraphics[width=0.32\textwidth]{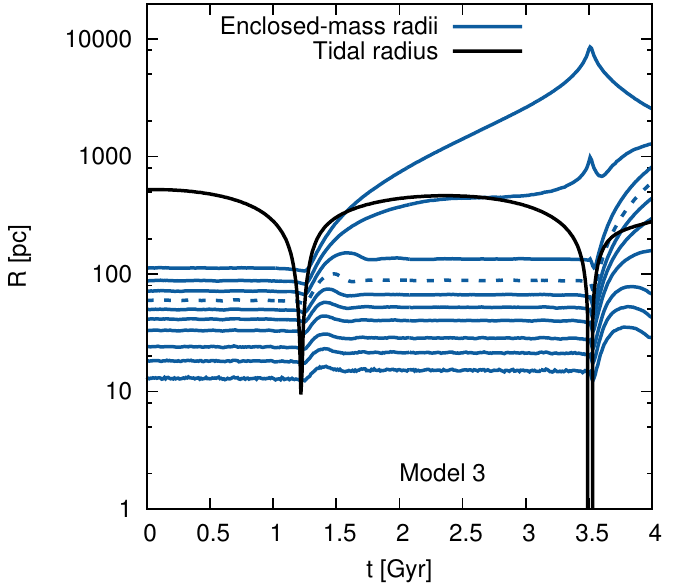}\\
\includegraphics[width=0.32\textwidth]{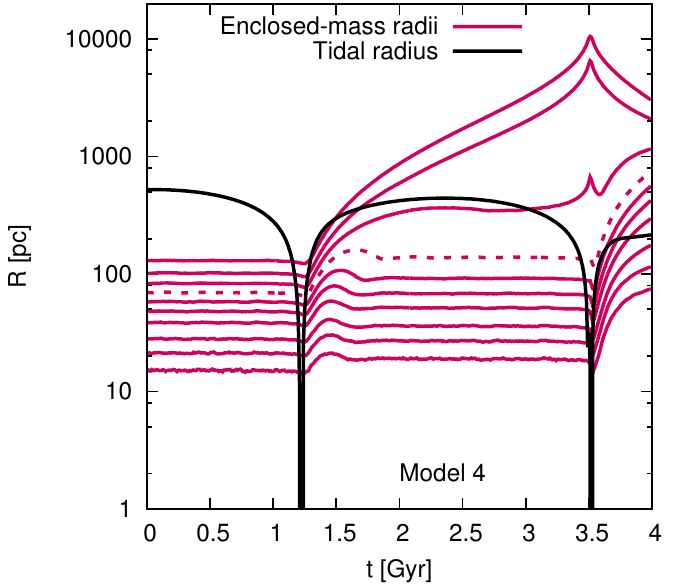}
\includegraphics[width=0.32\textwidth]{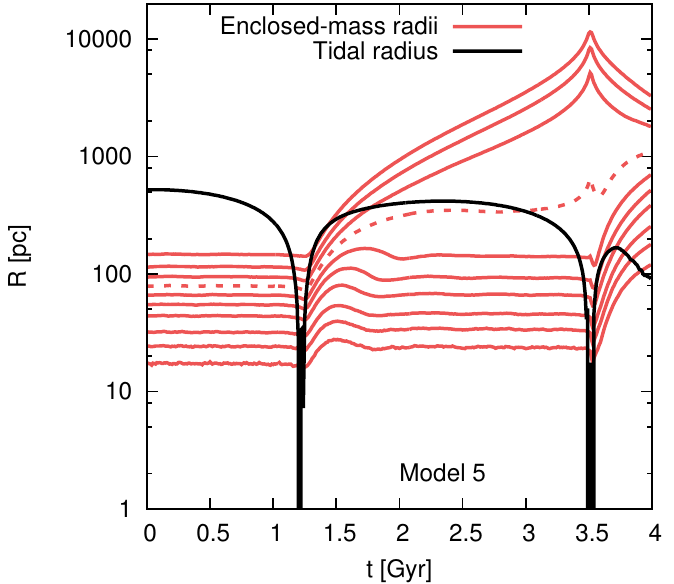}
\includegraphics[width=0.32\textwidth]{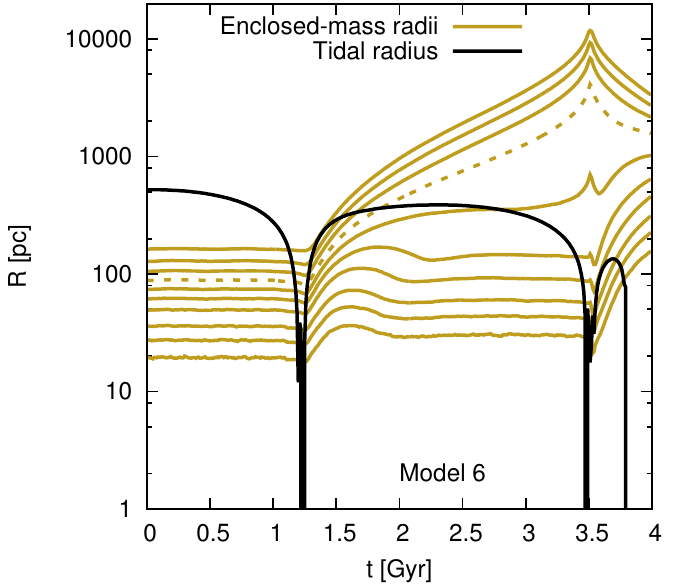}\\
\caption{Enclosed-mass radii, containing 2, 5, 10, 20 30, 40, 50 (dashed), 60, 70, and 80\% of the initial mass (from bottom to top). Also shown is the Jacobi radius, which cuts deep into the mass distribution during the two pericenter shocks, causing strong expansion of the mass shells. Expansion of the unbound mass shells reverses after pericenter, when the tidal debris is compressed back towards the progenitor, resulting in high debris densities during orbital phases in and around apocenter. Note that the 50\% enclosed-mass radius does not correspond to the half-mass radius. It is a measure of how much the initial configuration of the satellite spreads out over time.}
\label{KR}
\end{figure*}

Figure \ref{mass} shows the mass (left) and half-mass radius (right) evolution of the 6 $N$-body models. Model\,1 is the satellite with an initial half-mass radius of 40\,pc, while Model\,6 has an initial radius of 90\,pc. We determined the bound mass, $M_{bound}$, of the satellites by iterating the size of the instantaneous Jacobi radius \citep{King62}: 
\begin{equation}
r_J = \left( \frac{GM_{bound}}{\Omega^2-\partial^2\Phi/\partial R^2}\right)^{1/3},
\end{equation}
where $G$ is the gravitational constant, $\Omega$ is the satellite's angular velocity around the Galactic Center, and $\partial^2\Phi/\partial R^2$ is the second derivative of the Galactic potential with respect to the radius, $R$, at the Galactocentric distance of the satellite. To iterate $M_{bound}$, we first determine the density center of the satellite using a nearest neighbor scheme \citep{Casertano85}. We then sum up the masses of all stars within the Jacobi radius of the previous time step in the simulation. Based on this new mass, we calculate a new Jacobi radius estimate, and repeat this until the radius converges. 

As we can see in Fig.~\ref{mass}, the mass-loss of the satellites is strongly influenced by the tidal shocks at pericenter. The more extended the satellite is in the beginning, the more mass it loses during the two orbital periods. Model\,6 dissolves completely, Model\,5 barely survives with a bound core of $500\msun$, whereas the most compact model, Model\,1, loses merely 40\% of its mass. The mass loss is induced during pericenter passages, which are clearly visible in the bound-mass curves at 1.2\,Gyr and 3.5\,Gyr. During these passages, the tidal field is so strong that all models are temporarily unbound, meaning that the gravitational pull from the Galaxy is stronger than the internal gravitational forces within the satellites. However, the briefly unbound groups of stars move as ensembles out of pericenter and, although they are quickly spreading out, they can recapture into bound satellites. The bound mass in this newly formed satellite depends on how severely the satellite was affected by the pericenter shock. 

The same can be observed when looking at the half-mass radius of the bound mass (Fig.~\ref{KR}). Initially stable at the value that we set them up with, the half-mass radii of the models drop to zero during pericenter passages as the bound mass drops to zero. When they come out of pericenter, their half-mass radii grow rapidly. The more susceptible the satellite is to tidal shocking, the more its half-mass radius will grow through each shock. When a satellite suffers from severe mass loss through a shock, the half-mass radius grows temporarily to up to twice its initial size. As we will show below, this is due to an envelope of quickly expanding material that was blasted off the satellite through the pericenter shock.

\subsubsection{Tidal shocks causing explosive expansion}

Tidal shocking of the satellites has two consequences. First of all, the mass particles in the satellites gain energy, where the energy change per unit mass is proportional to
\begin{equation}
\left<\Delta E^2\right> \propto r^2v^2.
\end{equation} 
Here, $r$ is the radius of the particle from the center of the satellite, and $v$ is its velocity within the satellite (see, e.g., \citealt{Gnedin99}). This simple impulse approximation captures the most important behavior: the more extended the satellite is, the more it will be affected by tidal shocks. This energy input causes the satellites to expand. Secondly, the particles that are at larger satellite radii during the pericenter passage are more affected by the shocking. This causes a dispersion of the satellite particles, with the most affected ones speeding ahead or lagging behind. As a consequence of this, a fraction of the mass cannot be recaptured by the newly formed satellite on the way out to apocenter as the debris expands faster than the Jacobi radius can grow.

This behavior is demonstrated in Fig.~\ref{KR}, in which we show various enclosed-mass radii of the 6 models. These are the radii of mass shells around the satellite center, where we do not distinguish between bound and unbound particles. Also shown are the Jacobi radii (black solid lines), which cut deep into the mass shells during the pericenter passages. From the Jacobi radii it is obvious that Models\,1 and 2 do not become fully unbound during the first pericenter passage, but only at the second passage. 

Following the radii of the mass shells clearly demonstrates how the outer layers of the satellites get blasted off from the tidal shocks. The shells expand after each shock, and can only be recaptured by the Jacobi radius when the expansion of a shell is not explosive, that is, when the shell expansion rate is not equal to or larger than the expansion rate of the Jacobi radius.

\subsubsection{Orbital compression causing high surface densities}

Another important phenomenon can be observed in Fig.~\ref{KR}. The extent of the satellite and its debris is maximal in pericenter of its orbit. At 3.5\,Gyr, the debris is maximally stretched as the progenitor and its debris have reached their highest velocity in their orbit. After pericenter, the debris is compressed back into the vicinity of the satellite. This orbital compression can be substantial, reducing the size of the mass shells from 10\,kpc down to less than 5\,kpc. This causes expanding debris, which has just been blasted off the satellite and is expanding outward, to run into the compressing debris going inward,  yielding high densities well outside the Jacobi radius. For our Hercules model, the point where expanding and compressing debris meet, the ``debris pause'', is around 1--2\,kpc. In the case of Model\,6, more debris is outside this debris pause at the present day than inside.    

Thus, in our scenario, Hercules has lost large parts of its mass due to repeated, strong tidal shocks at pericenter, and only shows such a large, prominent appearance on the sky due to orbital compression of the debris. In the following section, we will have a look at the characteristics of our Hercules models in terms of observables.

\subsection{The structure of Hercules from N-body simulations}\label{sec:structure}

\begin{figure*}
\centering
\includegraphics[width=0.32\textwidth]{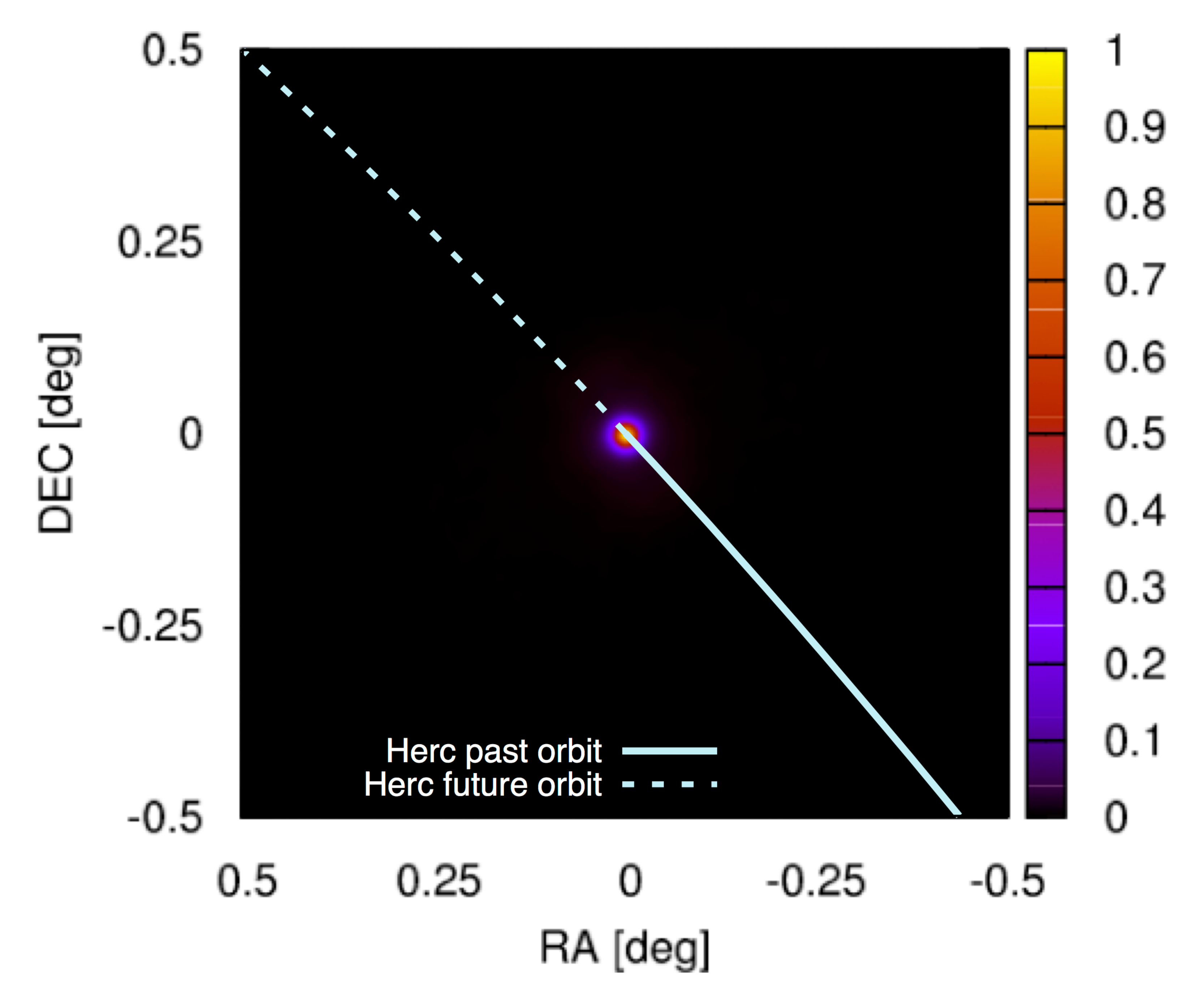}
\includegraphics[width=0.32\textwidth]{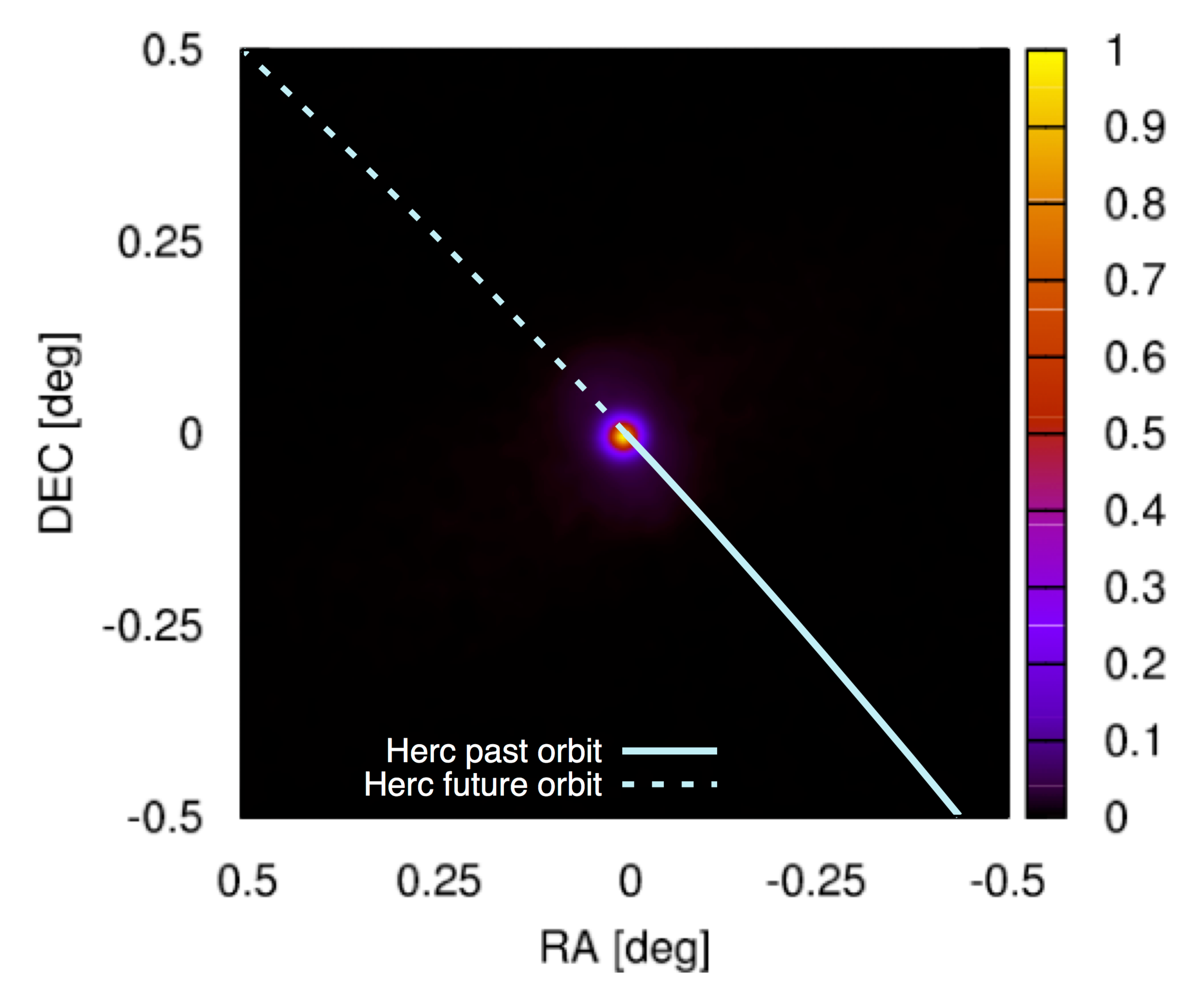}
\includegraphics[width=0.32\textwidth]{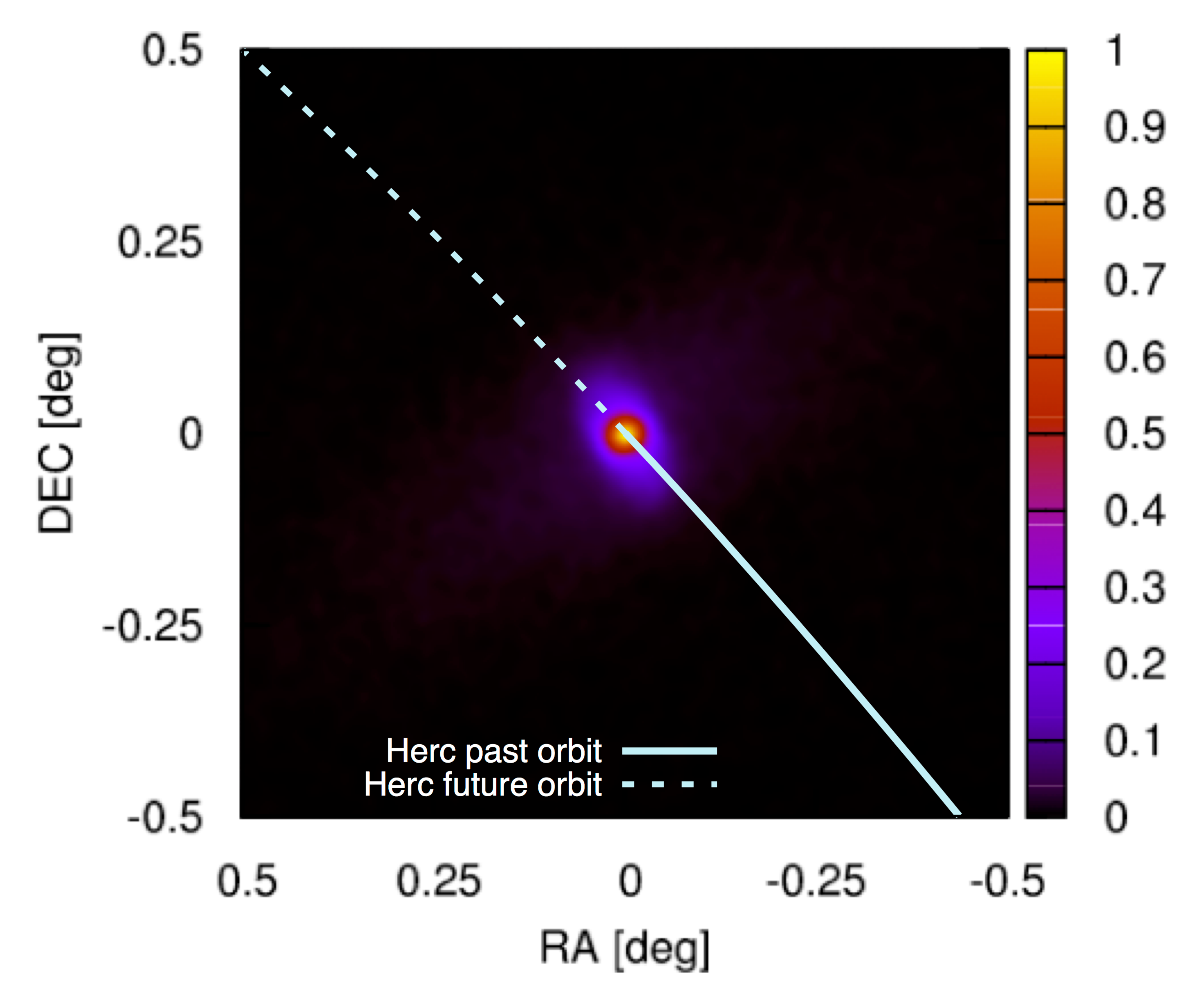}\\
\includegraphics[width=0.32\textwidth]{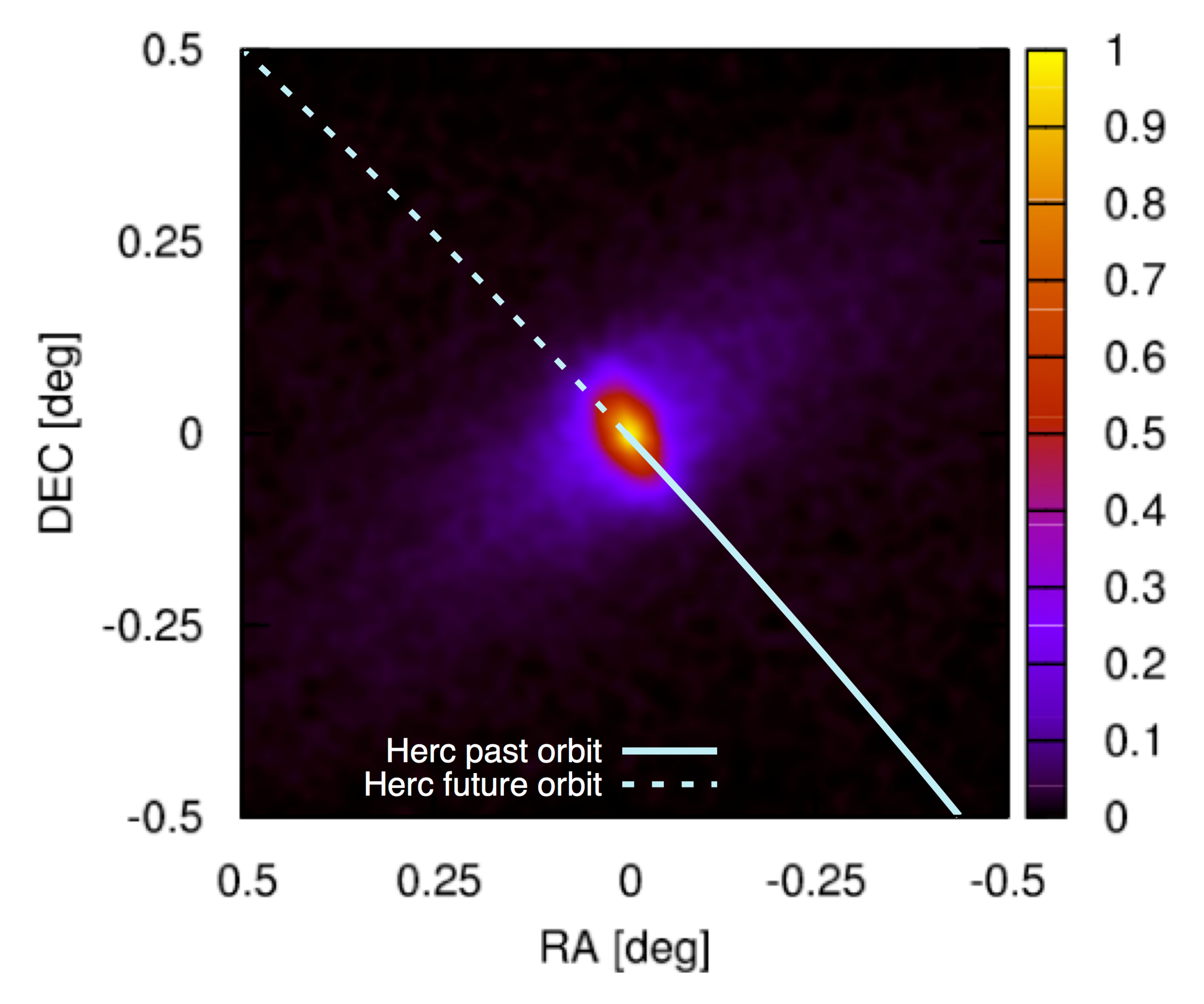}
\includegraphics[width=0.32\textwidth]{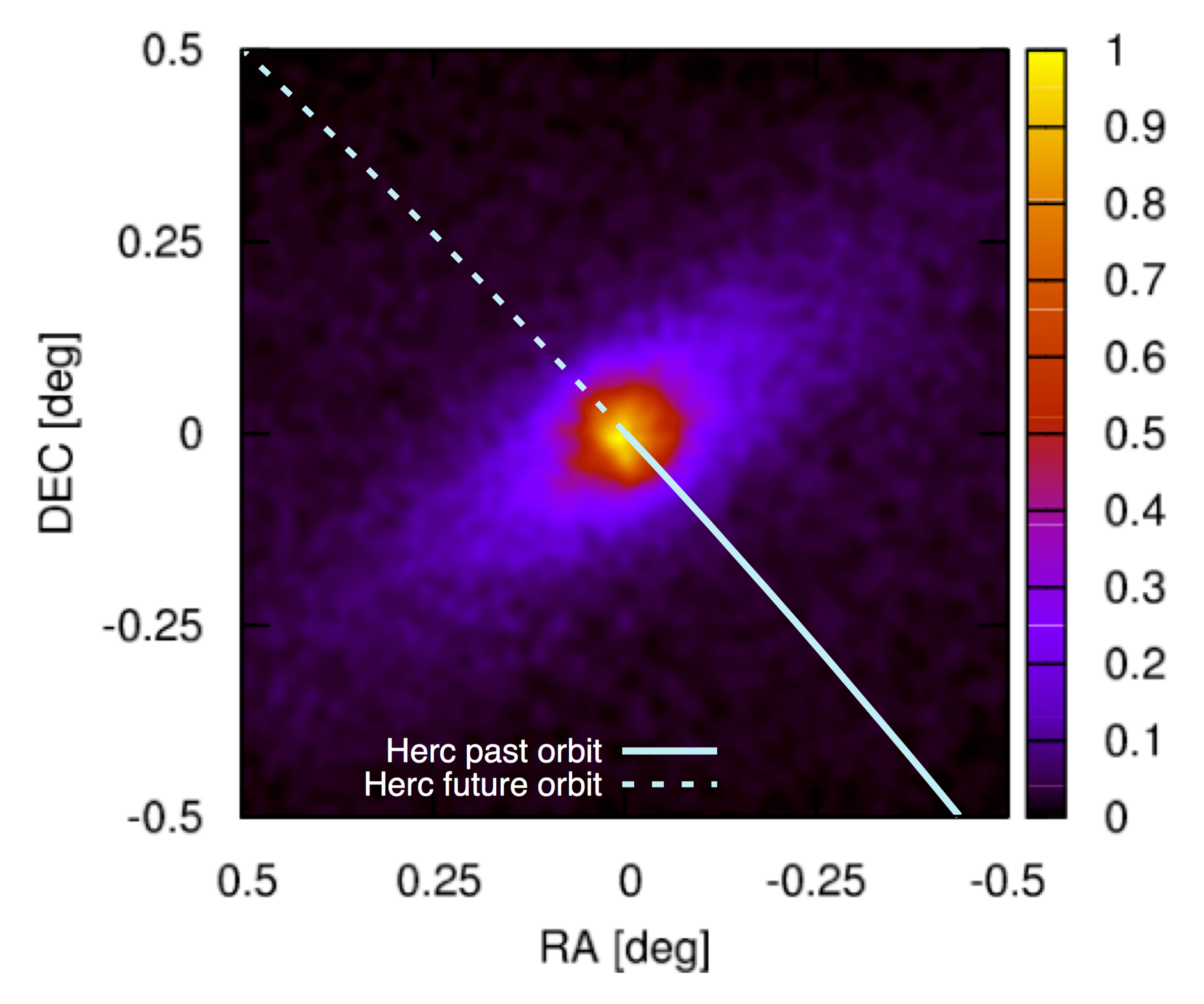}
\includegraphics[width=0.32\textwidth]{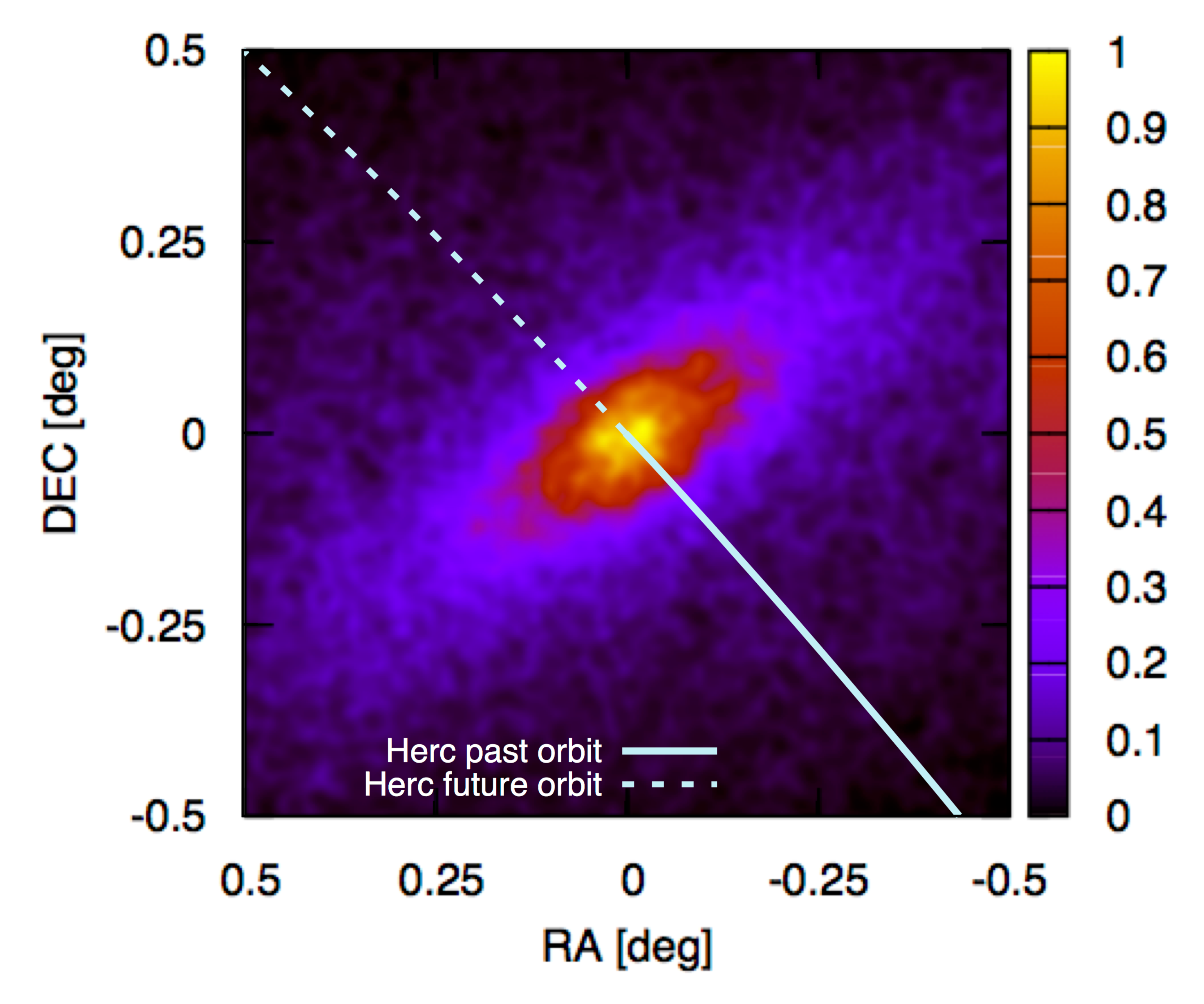}\\
\caption{Surface density distributions of the six Hercules simulations (Models\,1--6) as seen from the location of the Sun. The densities are normalized to the central surface density. The ``half-light'' density is therefore marked in red. The white line shows Hercules' orbit, and the dashed line its future orbit. The satellites show three distinct components: i) a round core, which disappears with more severe mass loss, ii) a tidal stream component along the orbit, which makes Hercules appear elongated along the orbit for Models\,3 and 4, and iii) the \textit{exploded component} perpendicular to the orbit, which gets pronounced for the nearly or completely dissolved satellites.}
\label{contours}
\end{figure*}

\begin{figure}
\centering
\includegraphics[width=0.45\textwidth]{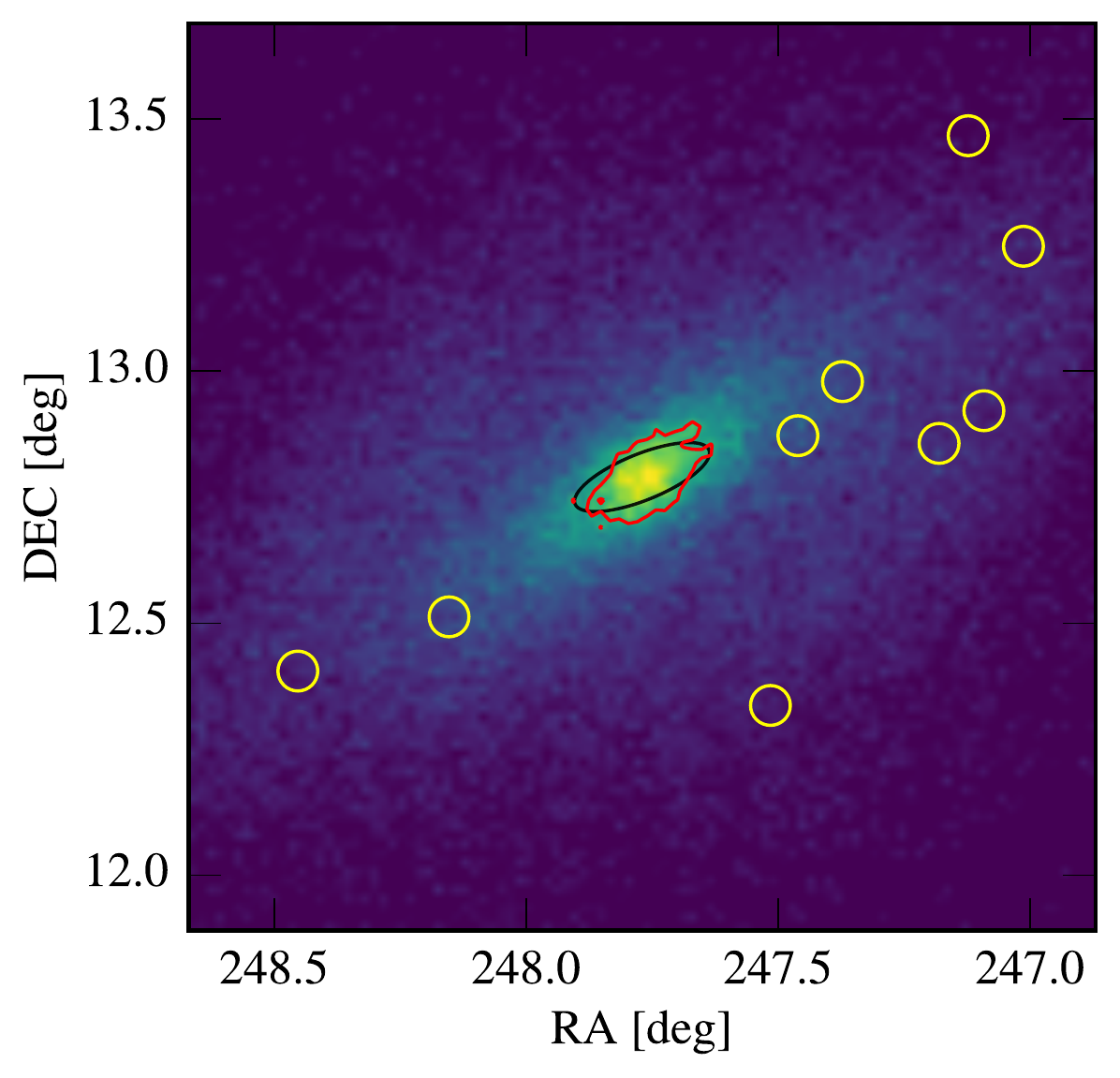}
\caption{Final snapshot of the fully dissolved Model\,6 as it would appear on the sky. Debris surrounds the satellite remnant in all directions, but preferentially along the SE/NW axis. Yellow circles mark the debris overdensities detected by \citet{Roderick15}. The half-light contour is highlighted by a red solid line. The irregular shape of the half-light contour is a clear sign of ongoing disruption. The black line indicates the elliptical contour fit to the satellite's half-light brightness profile from \citet{McConnachie12}. Without specifically fitting for it, the highly elliptic and irregularly shaped half-light contour of our model matches the data well. A comparison of the ellipticity of the models and the data can be found in Tab.~\ref{tab:profiles}. 
}
\label{halflight}
\end{figure}

\begin{table*}
 \centering
 \label{tab:profiles}
 \caption{Ellipse, 2D S\'ersic, and 2D Gauss fits to the $N$-body models}
 \begin{tabular}{cc|ccc|ccc|cc}
 Model & $r_{\rm eff,0}$ [pc] & $a$ [pc] &  $a$ [deg] & $\epsilon$ & S\'ersic $n$ & S\'ersic $r_{\rm eff}$ [deg] & S\'ersic $\epsilon$ & Gauss $\sigma_1$ [deg] & Gauss $\epsilon$\\
 \hline
1 & 30 & 29 & 0.012 & 0.05 & 1.1 & 0.013 & 0.09 & 0.010  & 0.05 \\
2 & 37 & 35 & 0.014 & 0.05 & 1.1 & 0.020 & 0.02 & 0.013  & 0.04 \\
3 & 45 & 53 & 0.022 & 0.11 & 1.5 & 0.047 & 0.13 & 0.027  & 0.15 \\
4 & 52 & 128 & 0.052 & 0.40 & 1.2 & 0.093 & 0.15 & 0.054  & 0.26 \\
5 & 60 & 142 & 0.058 & 0.03 & 1.4 & 0.219 & 0.39 & 0.108  & 0.39 \\
6 & 67 & 310 & 0.123 & 0.48 & 1.4 & 0.395 & 0.57 & 0.213  & 0.61 \\
\hline
\multicolumn{2}{c|}{\citet{McConnachie12}}  & $330^{+75}_{-52}$ & $0.143^{+0.030}_{-0.018}$ & $0.68\pm0.08$ & --- & --- & --- & ---  & --- \\

\end{tabular}
\end{table*}

Figure~\ref{contours} shows surface density maps of the 6 $N$-body simulations at the present day, as they appear when projected onto the sky and smoothed with a Gaussian kernel of 1\,arcmin width. The color coding is normalized to the central surface density. The ``effective contour'', where the surface density drops to half its central value, is colored in dark red. Also shown are the past (solid line) and future (dashed line) orbit of Hercules as they appear in projection.

The structures of the 6 models are very different. The sequence shows the transition from the ``bound'' (1, 2) to the ``tidal'' (3, 4) almost into the ``stream'' (5, 6) regime as described by \citet{Blana15}. The more the satellites suffered from the last tidal shock, the larger are their effective radii and the more elliptic and irregular they are.

Despite the recent tidal shocking, Model\,1 appears as a compact, circular satellite. Model\,2 is a bit more extended but also round and symmetric.  

Model\,3, which has lost 2/3 of its initial mass, shows signs of an elongated component outside its ``half-light'' radius of about 50--100\,pc, although its half-light contour is still fairly round. Its elongation is most pronounced within its instantaneous Jacobi radius, which measures 279\,pc or 0.1\,deg at this time. Comparing this panel of Model\,3 with the evolution of the mass shells in Fig.~\ref{KR} shows that the elongated component is formed by debris that has been affected by the last tidal shock and is now expanding rapidly. Model\,4 shows a significantly stronger elongation along the orbit with an ellipticity of 0.4. The entire satellite is in the process of expanding at the final snapshot.

Model 5 and 6 show increasingly larger present-day half-light radii, with Model\,6 being comparable to the real Herc's extent of 0.13\,deg, corresponding to 330\,pc. Unlike Model\,4, these two models have a strongly elongated debris component perpendicular to the orbit. Model\,5 appears to be simultaneously expanding along and perpendicular to the orbit, whereas Model\,6 is expanding primarily perpendicular to its orbit, matching the extent and orientation of the observed Herc debris very well (see also Fig.~\ref{halflight}). 

In order to measure the effective radii of the models and determine their ellipticity, we smooth the data with a Gaussian kernel using the SciPy/stats\footnote{http://www.scipy.org/} package \citep{Jones01}, where we fix the bandwidth to 1\,arcmin for better comparison among the models. We then fit an ellipse to the half-light contour using a python script\footnote{http://nicky.vanforeest.com/misc/fitEllipse/fitEllipse.html} that follows the algorithm outlined in \citet{Fitzgibbon96}. This method yields good fits even for Models\,5 and 6, which have irregularly shaped half-light contours  (Fig.~\ref{halflight}). The results are summarized in Tab.~\ref{tab:profiles} together with their initial effective radii, $r_{\rm eff,0}$, calculated using $r_h \approx 1.33\,r_{\rm eff}$ for a Plummer sphere \citep{Baumgardt10}. 

The half-light radii of the two compact Models\,1 and 2 do not expand throughout the simulations. But the models with higher susceptibility for tidal shocking expand significantly. Compared to their initial effective radii the semi-major axes of the final effective contour ellipses are up to 4 times larger. 

All models show somewhat elliptic half-light contours, a consequence of the recent tidal shocking. Models\,1 and 2 are close to spherical. Models\,3 and 4 have a significant elongation along the orbit, and Model\,6 has a strong elongation of 0.5 perpendicular to its orbit. Hercules' half-light ellipticity, for comparison, is about 0.65 \citep{Coleman07}. Model\,5 is a transition case between 4 and 6. It is capturing the moment when the orientation of the debris ``flips'', as described by \citet{Blana15}.

We also fit two parametric models to the data, a two-dimensional S\'ersic model and a two-dimensional Gaussian. We use the Astropy/modeling package for this task \citep{Astropy13}. Due to their different profile slopes, the two parametrizations capture different aspects of the six models. Both tend to be more extended than the non-parametric half-light measurement, which is due to a bad fit to the data in the center of the satellite. But the trends in effective radius and ellipticity are basically the same for all three methods.

Figure~\ref{halflight} shows that Model\,6 captures some important aspects of the real Hercules. It is strongly elongated with almost the same orientation as Hercules (cf., half-light contour ellipse from \citealt{McConnachie12}), without us specifically fitting our model to this ellipse. At the same time Model\,6 is vastly extended, and has a low total (stellar) mass. The positions of the over-dense regions observed by \citet{Roderick15} are shown as yellow circles.

The questions we will answer next are: where exactly does its elongation come from and why is it perpendicular to the orbit? As we will see in the following section, the large extent, the low mass, and the orientation of the elliptical mass distribution perpendicular to the orbit are all a consequence of differential orbital precession.

\section{Stream fanning from differential orbital precession}\label{sec:differential}

\begin{figure}
\centering
\includegraphics[width=0.45\textwidth]{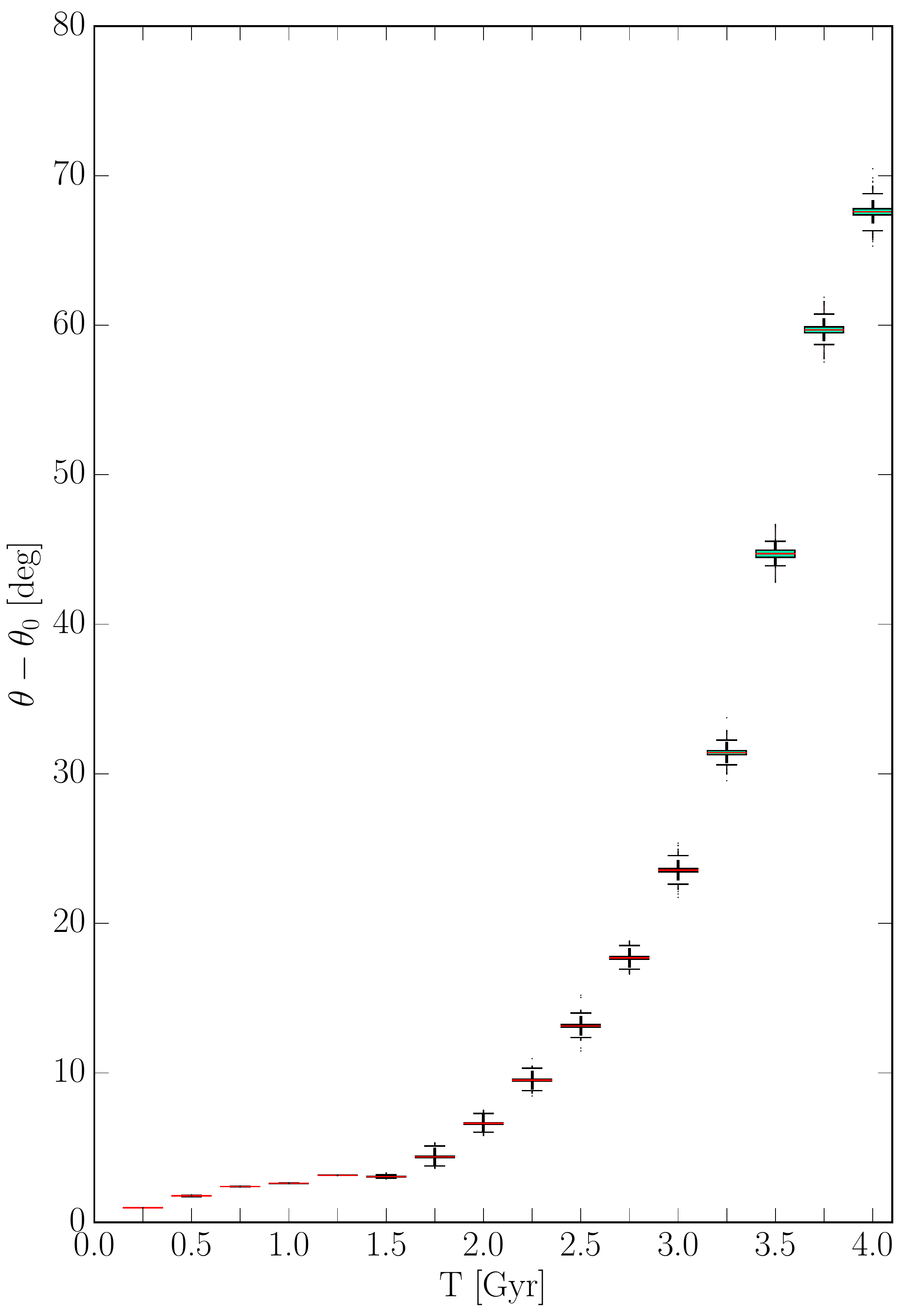}
\caption{Box-and-whisker plots showing the orientation angles of orbital planes for particles in an idealized streakline simulation. The test satellite orbits on our best-fit orbit for Herc in the respective oblate galactic potential. Initially at 0\,Gyr, all test particles have the same orbital plane orientation, $\theta_0$, but with time the orientation angles change as the whole satellite precesses due to tidal torques from the galactic potential (the red line shows the median or the orientation angles). Differential orbital precession causes the test particles to spread in orbital plane orientation (indicated by the boxes and whiskers), which is most effective during pericenter passages at 1.2\,Gyr and 3.5\,Gyr, when the satellite orbits through the inner 5\,kpc of the Galaxy.}
\label{precession}
\end{figure}

\begin{figure}
\centering
\includegraphics[width=0.45\textwidth]{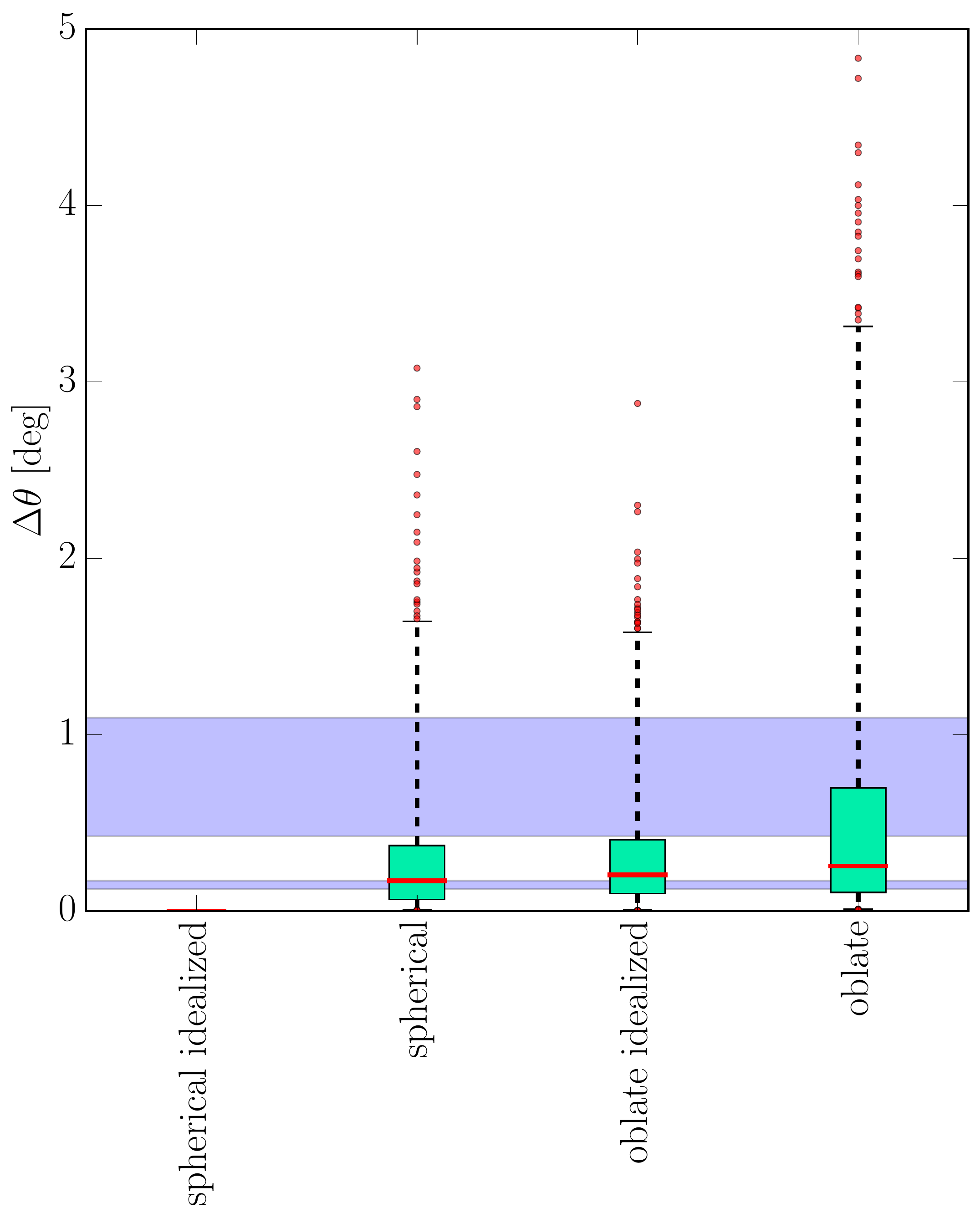}
\caption{Distributions of difference angles in orbital plane orientations, $\Delta\theta = \theta - \theta_{median}$, of particles from four different streakline models after 4\,Gyr of integration. The red lines show the medians of these distributions, the boxes indicate the lower to upper quartiles, and the whiskers mark the upper and lower 2.5\% of the distributions. Fliers show the most extreme points. The \textit{spherical idealized} case has no scatter in escape conditions, hence all particles of this model have the same orbital plane and keep it throughout the simulation. Adding scatter in escape conditions perpendicular to the orbital plane introduces significant differences in orbital plane orientations (\textit{spherical}). The same order of effect can be achieved from differential orbital precession in an oblate potential (\textit{oblate idealized}), and the spread is doubled when both effects are combined (\textit{oblate}). In the latter case, 75\% of particles have inclinations of up to 0.7\,deg with respect to the center of the satellite, corresponding to 1.7\,kpc at the distance of Hercules. The lower horizontal bar marks the effective radius of Hercules (0.12\,deg), and the upper bar indicates the mean distance (and dispersion) of the overdensities from the center of Hercules (see Tab.~\ref{tab:overdensities}).
}
\label{precession_comparison}
\end{figure}

\citet{Blana15} observe a flip in the orientation of the debris for simulations of Plummer spheres with $10^5\msun$ and half-mass radii of 100\,pc over 5\,Gyr, which is similar to our configurations. However, their orbit was chosen such that it aligns with the elongation of the observed Hercules \citep{Martin10}. Therefore, the flip misaligns the elongation with the observations. In our case, however, we chose the orbit such that the flipped debris would reproduce the observed debris overdensities around Hercules \citep{Roderick15}. How did our streakline models know the debris would flip eventually?

The flipping is in fact a fanning \citep{Pearson15}: when we look at the orbital poles (the direction of the angular momentum) of the satellite and each of its debris stars we can see that they spread significantly throughout the simulation.  Figure~\ref{precession} shows a number of box-and-whisker plots for an idealized streakline model at different timesteps during the past 4\,Gyr. The satellite's initial angular momentum vector defines the reference orbital pole, $\theta_0$, measured with respect to Galactic north. After two orbits the difference between orbital plane orientations among the debris stars and the satellite has grown to several degrees.

In an idealized Galaxy (spherical), an idealized satellite (producing zero-temperature ejecta) would not show such a fanned debris distribution. Two factors, non-zero velocities of stream stars perpendicular to the orbital plane, and a non-spherical galactic potential, can cause this fanning:
\begin{enumerate}
\item The satellite ejects stars (or stars are stripped from the satellite) on orbits with slight intrinsic differences in the orbital planes to the satellite's orbital plane. That is, escaping stars can have offset velocities perpendicular to the orbital plane of the satellite. This effect is enhanced by severe tidal shocks  as the amount of random motion is higher for such violently stripped stars (see, e.g., \citealt{Hendel15}).    
\item The non-spherical components of the Galactic potential can exert a torque on the orbital plane of the satellite and its debris. Depending on the torque angle and its leverage, the degree of precession can vary significantly across the extent of the debris (see, e.g.,  \citealt{Erkal16}).
\end{enumerate}
We show the first effect for our Hercules orbit in Fig.~\ref{precession_comparison}, which we measured by using a Galactic model with a spherical dark matter halo, and no disk component. Due to spherical symmetry, there is no precession in this potential so that satellite and debris stars keep orbiting in the same plane (case \textit{spherical idealized}). 

In the same spherical potential we also created a streakline model with significant random scatter in the offset positions and velocities around the Lagrange points (case \textit{spherical}). The velocity offsets were generated using a Gaussian distribution with a width of 2 km\,s$^{-1}$ in each velocity component. For a satellite on a Hercules-like orbit, the effect can produce difference angles in orbital poles of about one degree.  

We estimate the strength of the second effect by setting up perfectly cold streams in our oblate Galaxy potential, meaning that the streakline particles are released in the instantaneous orbital plane of the progenitor and have the same angular velocity as the satellite but no velocity perpendicular to the orbital plane (case \textit{oblate idealized} in Fig.~\ref{precession_comparison}). Both effects are of the same order of magnitude. Depending on the type of satellite and orbit one or the other effect may be more important. In fact, differential orbital precession may be a viable explanation for the stream fanning observed by \citet{Pearson15} for the Palomar\,5 stream in their triaxial trial potential.

Satellites on Hercules-like orbits will experience a combination of both effects, resulting in a spread in orbital pole orientations that is nearly twice as large as each individual effect (\textit{oblate}). In this case, the median orbital plane difference angle is 0.25\,deg, and about 75\% of all streakline particles have inclinations of up to 0.7\,deg with respect to the center of the satellite after just two orbits around the Galactic Center. This difference in orbital planes corresponds to a fanning of 1.7\,kpc at the heliocentric distance of Hercules of 140\,kpc. Since the difference angle can be positive and negative, the spread in orbital planes gives the satellite and its debris an extent of more than 3\,kpc perpendicular to its orbit. 

Although our choices for the number of particles in our streakline models was rather arbitrary, we show how this spread compares to the observed effective radius of Hercules and the tidal features detected by \citet{Roderick15} in Fig.~\ref{precession_comparison}. The authors report that their detected overdensities contain at least as many stars as the main body of Herc. Only our \textit{oblate} case would be able to explain such a strong spread in orbital plane orientations.  

As we will show in the following section, the length of the satellite and most of its debris along the orbit is less than 1\,kpc. The visible part of the Hercules stream is therefore wider than it is long. This unexpected shape may, in fact, be used to constrain Hercules' nature and its orbit.

\section{Signatures of an exploding satellite and future observations}\label{sec:signatures}

\begin{figure*}
\centering
\includegraphics[width=0.90\textwidth]{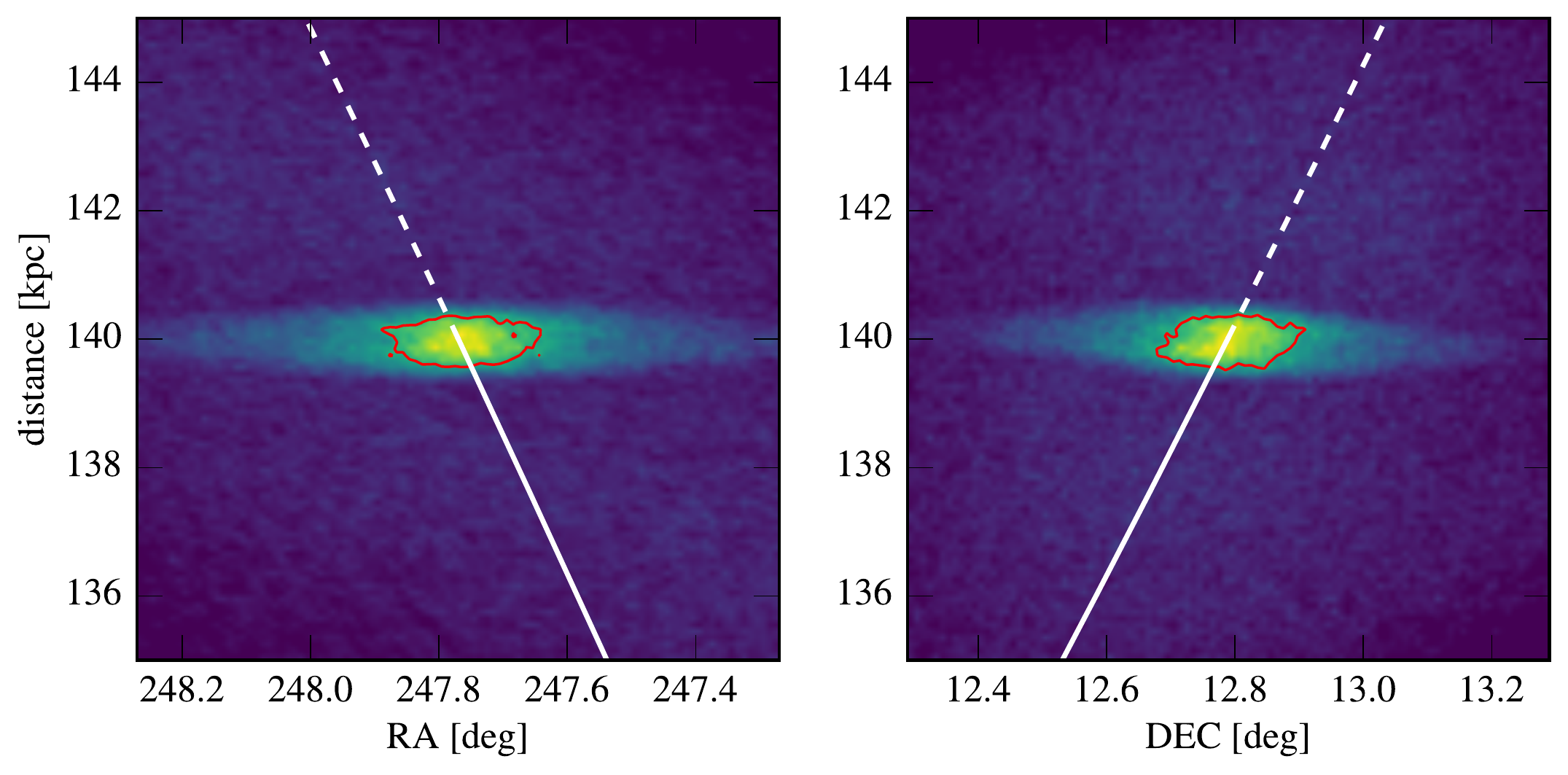}
\caption{Heliocentric distance gradient of particles along the satellite and its stream -- shown here for the completely dissolved Model\,6. The (barely visible) subdominant stream component loosely follows the distance gradient of the satellite orbit (indicated by white solid and dashed lines for the past and future orbital path, respectively), whereas the \textit{exploded component} has not spread along the orbit yet and hence shares the same distance as the progenitor to within $\pm0.4$\,kpc (the red contour indicates where the density drops to half its central value). Such a distance gradient (or lack thereof) could be detected with deep photometric data by making differential distance measurements along the extent of the satellite.}
\label{distance}
\end{figure*}

\begin{figure*}
\centering
\includegraphics[width=0.90\textwidth]{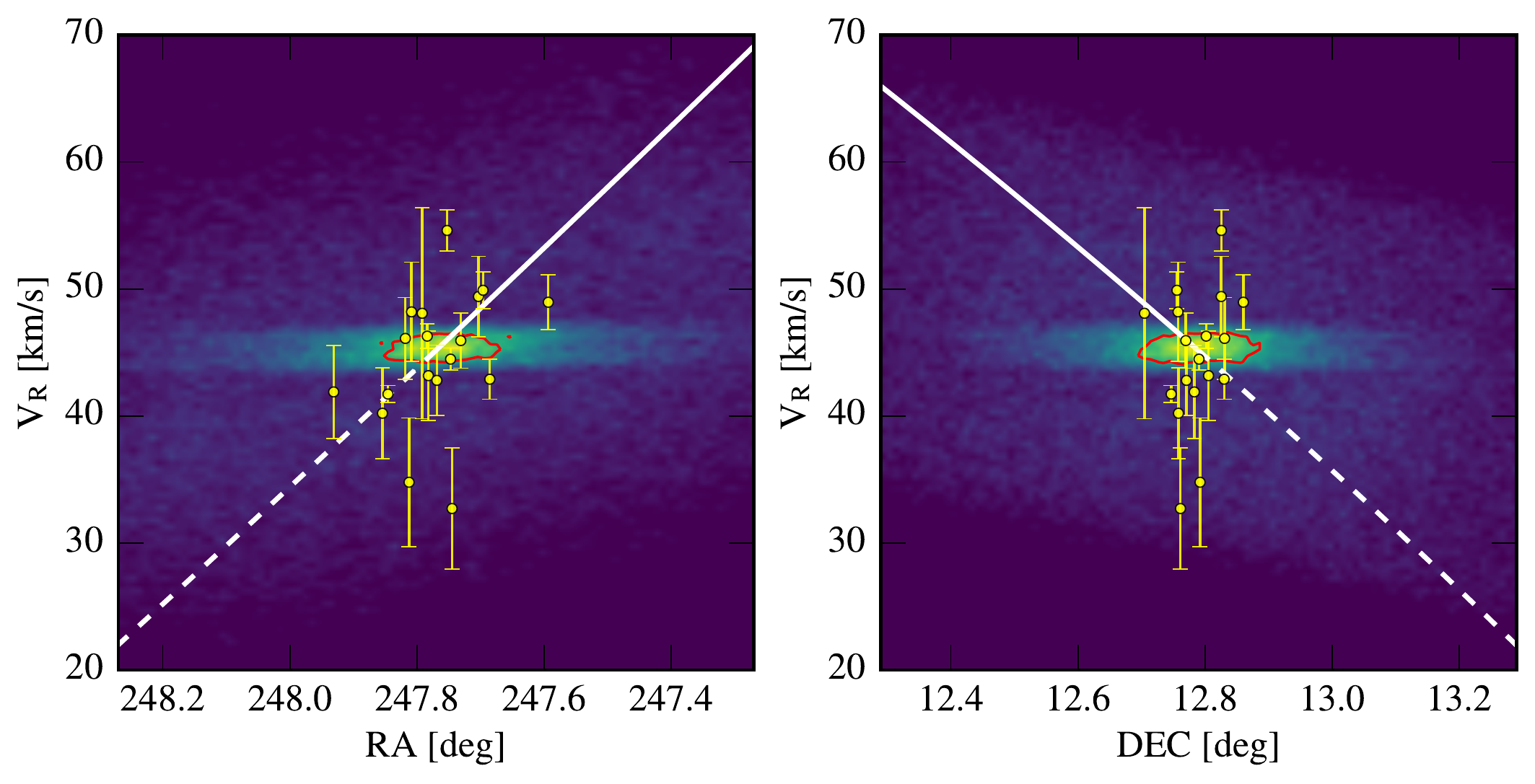}
\caption{Radial velocity distribution of particles from the dissolved Model\,6. As for the distances of particles in Fig.~\ref{distance}, the faint, subdominant stream component roughly follows the radial velocity gradient of the satellite orbit (white line), whereas the \textit{exploded component} has the same radial velocity as the satellite remnant to within $\pm 1.2$\,km\,s$^{-1}$. The yellow markers are the radial velocity measurements from \citet{Aden09b}. The \textit{exploded component} of Hercules forms a distinct velocity substructure and should be detectable with sufficiently precise data (see also Fig.~\ref{radialvelocity_histogram}). Available spectroscopic samples may be biased towards stars from the trailing part of the stream, which is at smaller heliocentric distances and has a larger radial velocity (cf.~Fig.~\ref{distance}).}
\label{radialvelocities}
\end{figure*}

\begin{figure}
\centering
\includegraphics[width=0.45\textwidth]{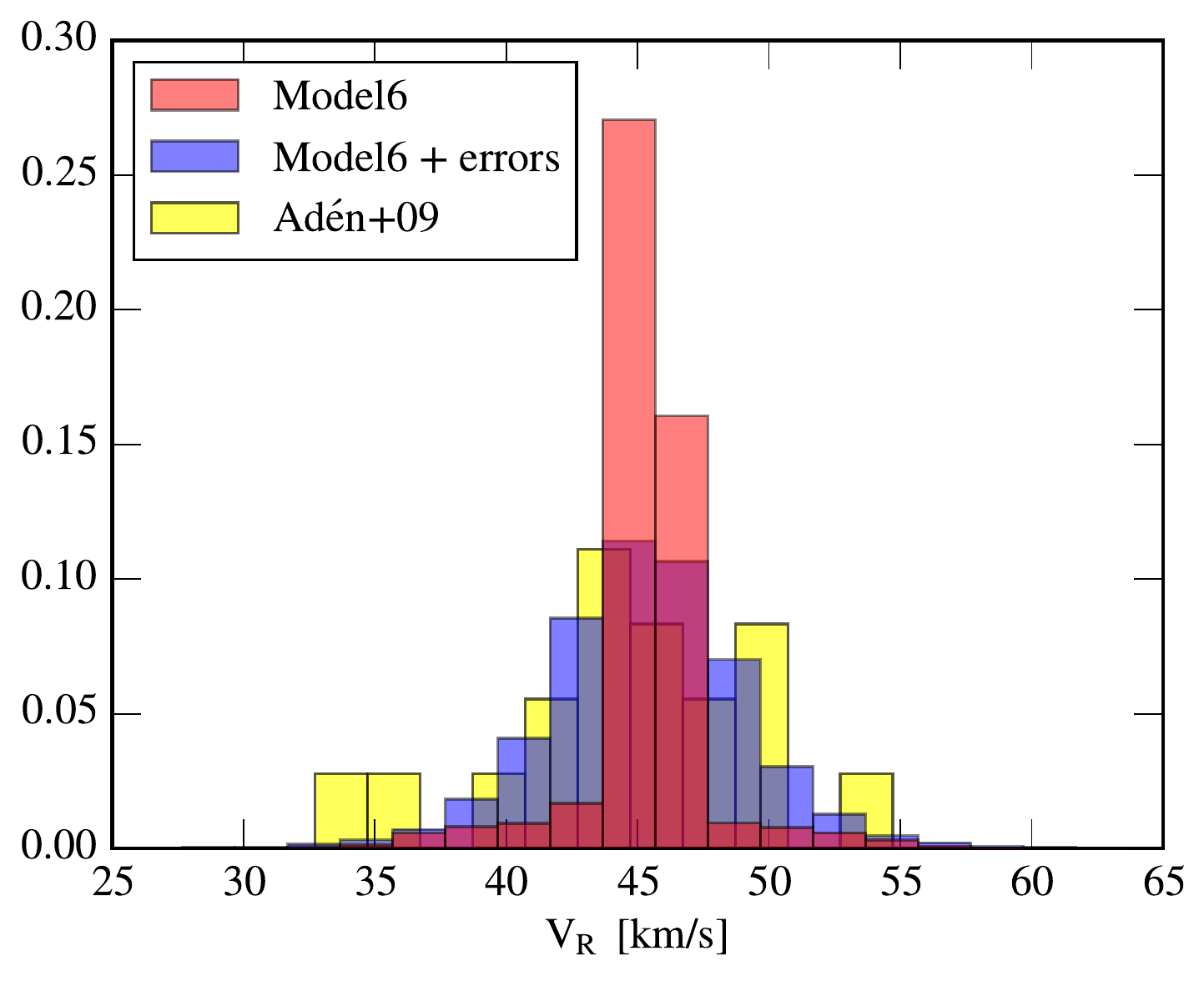}
\caption{Normalized histograms of radial velocities from \citet{Aden09b}, our $N$-body Model\,6, and Model\,6 with added Gaussian random errors of mean size 3\,km\,s$^{-1}$. The $N$-body model shows a strong peak at the systemic velocity of the progenitor, corresponding to the \textit{exploded component}.  Uncertainties like the ones of the \citet{Aden09b} sample blur our this peaked distribution since the peak width of $\pm1.2$\,km\,s$^{-1}$ is smaller than the average per-star uncertainty in the Ad\'{e}n sample.}
\label{radialvelocity_histogram}
\end{figure}

A low-density satellite, like our simulated Hercules, that got disrupted in only two revolutions about the Galactic center has a quite short debris stream (in terms of length along the orbit compared to length of one orbit). Following \citet{Kupper10}, we can estimate the length of the stream for the hypothetical case that our satellite was on a circular orbit at the current galactocentric distance of Hercules. Assuming a mass of about 50\,000\msun\,and a circular velocity of about $V_C = 200$\,km\,s$^{-1}$ at 140\,kpc, the instantaneous tidal radius of Hercules would be
\begin{equation}
r_t = \left( \frac{GM}{2\Omega^2}\right)^{1/3}\approx 380\,\mbox{pc},
\end{equation}
where $\Omega$ is the satellite's angular velocity. The mean drift velocity of stars along the orbit of Herc (with respect to the satellite) would then be 
\begin{equation}
\overline{v}_{drift} \approx \pm2\Omega r_t = \pm1.1\,\mbox{pc}\,\mbox{Myr}^{-1}. 
\end{equation}
Within our simulation time of 4\,Gyr, the bulk of the stream would have grown to no more than about $\pm4$\,kpc length along the orbit (in each direction). But since Hercules is on a highly eccentric orbit and is currently approaching apogalacticon at a velocity of  $V=157$\,km\,s$^{-1}$ (see Sec.~\ref{sec:streakline}), its debris is significantly compressed (approximately by a factor of $1-V/V_C\approx 1/4$) as the whole debris-satellite system is decelerating. Hence it will be shorter than 4\,kpc, and will keep being compressed until it reaches apogalacticon, where the satellite's velocity is as low as 23\,km\,s$^{-1}$. A difference angle in orbital poles of about $\pm0.7$\,deg among debris stars (see Fig.~\ref{precession_comparison}) is therefore substantial for such a system, since, at the heliocentric distance of 140\,kpc, this difference in orbital poles corresponds to a physical (projected) separation of $\pm1.7$\,kpc, which is of the order of the compressed stream's length along the orbit.  

Furthermore, in our simulations, Herc has lost most of its mass during the last pericenter passage 500\,Myr ago. Since then, the debris has spread no more than $\pm500$\,pc along the orbit. As we will show in the following section, this part of the debris, the \textit{exploded component}, has about the same angular extent as the whole stream, which makes this part of Herc wider than long.

\subsection{Observational tests}
This unexpected case of an elongated satellite that is moving perpendicular to its apparent major axis has two major consequences for observations of Hercules. First of all, there should be no strong distance gradient along the extent of the youngest part of the stream, which was formed during the last tidal shock, that is, along the \textit{exploded component}. Secondly, radial velocity measurements are likely to pick up on different components of the satellite-stream system and get misleading results when assuming virial equilibrium. Both signatures will be described in more detail below.

In Fig.~\ref{distance}, we show the heliocentric distances of particles from our 4\,Gyr-long $N$-body simulation of Hercules (Model\,6) at the final snapshot. There is a dominant component at the distance of Herc that is spread out in right ascension and declination but does not follow the track of Herc's orbit. This is the \textit{exploded component} that originates from the tidal shock Herc experienced during the last pericenter passage 500\,Myr ago. We measure the width of this component by fitting an ellipse to the contour at which the density drops to half its central value. The half-width of this ellipse in the distance dimension is 0.42\,kpc. 

Since the \textit{exploded component} dominates the stellar profile of Herc, photometric distance measurements should not see any distance gradient along this part. In fact, existing deep HST photometry of Hercules' center \citep{Brown14} could be leveraged with new, equally deep HST photometry of off-center fields along the stream to measure high-precision, differential distances between the center of Herc and the \textit{exploded component}.

Figure~\ref{radialvelocities} shows the radial velocities of the same data as in Fig.~\ref{distance}. The \textit{exploded component} dominates the radial velocity profile of Hercules. It has a width of only $\pm1.2$\,km\,s$^{-1}$ even though the satellite is completely unbound at this stage, and this velocity is nearly constant across the whole component. In contrast to that, the subdominant stream component loosely follows the radial velocity gradient of the satellite's orbit (white line), which is comparable to the tentative $16\pm3$\,km\,s$^{-1}$ gradient measured by \citet{Aden09a}. This should come at no surprise since we used the radial velocities from \citet{Aden09b} as input data for our Bayesian modeling (shown as yellow data points).

But there is more information in the available spectroscopic data: In their sample of 30 spectroscopic measurements of likely Herc members, \citet{Simon07} find an overdensity of 9 stars between 41 and 43\,km\,$^{-1}$. They conclude that such a velocity substructure is rather unlikely and may be indicative of ongoing tidal disruption. Given our scenario of a strongly gravitationally shocked Hercules, the overdensity found by \citet{Simon07} may very well correspond to the \textit{exploded component} of Herc. The other 21 candidates would then correspond to stars in the stream component or foreground contaminations \citep{Aden09b}. 

The distribution of spectroscopic targets in the \citet{Simon07} data is asymmetric around this velocity substructure. Our orbit prediction would suggest that this is due to the strong distance gradient of the stream component. The spectroscopic sample would then be biased towards the trailing tail (solid white line in Figs.~\ref{distance} \& \ref{radialvelocities}), which lies at smaller heliocentric distances.

Such a velocity substructure is not visible in the \citet{Aden09b} data. To check if it should be detectable, we selected particles from Model\,6 in the same sky region as the Ad\'en spectroscopic sample. A normalized histogram of the $10^4$ radial velocities in this sky region is shown in Fig.~\ref{radialvelocity_histogram}. With perfect data, the velocity substructure should be clearly visible. The Ad\'en sample has an average per-star uncertainty of 2.97\,km\,s$^{-1}$ (comparable to the average uncertainty in the \citealt{Simon07} data). If we add Gaussian random errors with this average size to the model points, the distribution matches the width of the observed distribution. If we draw 18 particles from this sample and calculate the velocity dispersion of this sample, and repeat this 1000 times, we get a mean value of $3.5$\,km\,s$^{-1}$ with a standard deviation of 0.7\,km\,s$^{-1}$, which compares very well to the $(3.7\pm0.9)$\,km\,s$^{-1}$ reported by \citet{Aden09b}. Future spectroscopic data with higher precision should be able to clearly resolve this velocity substructure.  

At this point it is important to remember that our $N$-body simulation covered only the past 4\,Gyr. A satellite that fell into the gravitational potential of the Milky Way at earlier times will have a much more extended, and probably more pronounced, stream component. Moreover, we have to keep in mind that we fixed the Galactic gravitational potential to the one determined in \citet{Kupper15}. Since Hercules is so far away from the Galactic center, the geometry of Hercules is very fine-tuned. Hence, a small change in the choice of the potential will have a strong influence on the exact velocity structure. Gathering observational constraints on the dynamical state of Hercules will therefore be extremely valuable for constraining the shape of the Galactic potential.

\section{Conclusions}\label{sec:conclusions}

Using streakline modeling and $N$-body simulations, we have shown that the published observational data on the Milky Way ultra-faint satellite Hercules can be explained by a disrupting low-mass satellite on a very eccentric orbit with $\epsilon\approx0.95$ (Fig.~\ref{orbit}). We limit our investigation to the formation of the extended, elongated main body of Hercules as seen in recent SDSS, LBT and DECam data. We find that this part of Hercules may have formed during the last pericenter passage as a consequence of strong tidal shocking. For this reason, we limit our investigation to the last 4\,Gyr, corresponding to two full orbits about the Galaxy, without investigating further the nature or origin of Hercules. 

In our scenario, the satellite experiences strong, episodic tidal shocks at pericenter of its orbit, ``explosively'' removing large fractions of its mass (Fig.~\ref{KR}). On our best estimate for Hercules' orbit, the satellite is currently approaching apocenter at 185\,kpc Galactocentric distance, after experiencing a disruptive tidal shock at 5\,kpc perigalactic radius 500\,Myr ago. From $N$-body simulations we estimate that the satellite must have had a very low central density ($<0.01\msun\,$pc$^{-3}$) before experiencing this last tidal shock (Fig.~\ref{contours}), and must have had a low central density of $\approx0.01\msun\,$pc$^{-3}$ already 4\,Gyr ago. It would be interesting to see if disrupting satellites with such low central densities are a natural outcome of structure formation in a $\Lambda$CDM universe.

We conclude that the object formerly known as ultra-faint dwarf galaxy Hercules is at the present day most likely a largely (or completely) unbound structure. Given these model constraints, we conclude that Hercules is unlikely to presently contain a high dark matter concentration and may therefore not be a promising candidate for dark matter search experiments. In fact, we do not require Hercules to have a dark matter component in our scenario, and only the large spread in stellar metallicities among the Hercules stars is indicative of Hercules having been a galaxy once. This is in tension with simulations showing that nearly disrupted dwarf galaxies with cuspy dark-matter halos are strongly dark-matter dominated \citep{Penarrubia08, Penarrubia10}.

We find that the reason for Hercules' particular shape at its current location 140\,kpc from the Galaxy is differential orbital precession of its stars as they were deflected in the aspherical gravitational potential of the Milky Way (Fig.~\ref{precession}). For our assumed galaxy potential, the difference angle between orbital plane orientations of different parts of the stream is $>1$\,deg at the present day (Fig.~\ref{precession_comparison}), which corresponds to more than 2\,kpc angular extent on the sky at the distance of Hercules. Due to this effect, the distribution of Hercules stars (bound or unbound) is significantly larger perpendicular to the orbit than it is along the orbit, making Hercules the only known stream that is broader than it is long. 

Our scenario makes testable predictions: The last pericenter shock should have induced significant mass loss, and Hercules should therefore show a prominent feature from this shock in the form of expanding tidal debris. We predict this \textit{exploded component} to extent (at least) out to Hercules' effective radius of $\approx 330$\,pc, and may likely be the origin of all debris structures found by \citet{Roderick15} out to 1.3\,deg from the center of the satellite. This component should not show any distance gradient (Fig.~\ref{distance}) or radial velocity gradient (Fig.~\ref{radialvelocities}), but with given spectroscopic precision should be clearly distinguishable in radial-velocity space from the less pronounced stream component (Fig.~\ref{radialvelocity_histogram}). 

The unique geometry of the Hercules tidal debris and its exceptional distance from the Sun make our predictions for Hercules' proper motions very precise ($\mu_{\alpha} \cos(\delta) = -0.210^{+0.019}_{-0.013}\,\mbox{mas}\,\mbox{yr}^{-1}$, $\mu_\delta = -0.224^{+0.015}_{-0.016}\,\mbox{mas}\,\mbox{yr}^{-1}$). These proper motions correspond to velocity components as low as $139^{+13}_{-9}$\,km\,s$^{-1}$ and $149^{+10}_{-11}$\,km\,s$^{-1}$. Together with its receding radial velocity of 45.2\,km\,s$^{-1}$ and put into the Galactic rest-frame, Hercules' total velocity adds up to no more than $157$\,km\,s$^{-1}$ in our scenario, of which 99\% is directed radially outwards. 

However, this is only the prediction for the given Galactic potential that we tested here \citep{Kupper15}. Changing the shape of the Galaxy model would change this prediction. Vice versa, high-precision measurements of Hercules' true proper motions with, e.g., Gaia will therefore be able to put strong constraints on the potential of the Milky Way. Yet, Gaia will be able to measure proper motions only for the brightest stars in Hercules \citep{deBruijne14}, which are the probable G and K giants from \citet{Aden09a}. At the distance of Hercules, these RGB stars have V magnitudes of about 20\,mag, resulting in Gaia measurement uncertainties of $\approx$0.2\,mas\,yr$^{-1}$, that is, an uncertainty in the tangential velocity of about $130$\,km\,s$^{-1}$. However, leveraging Gaia data with archival HST data, or waiting for LSST and especially WFIRST to come online, will enable us to put significantly stronger constraints on the proper motion of Hercules. 

But proper motion is just one way to test this scenario. Measurement of a distance gradient (or a lack thereof) across the body of Hercules is already possible with deep HST imaging. Given enough of such observational constraints, Herc may, in fact, be used as one of the most powerful probes of the shape of the Galactic gravitational potential. 

Other UFDs with signs of ongoing tidal disruption may also be expanding in response to intense gravitational shocks. Hence, exploding satellites may pose a new powerful family of galactic potential tracers.

\section*{Acknowledgements}

The authors would like to thank Tammy Roderick for providing the coordinates of their identified tidal features. The authors also thank Sverre Aarseth and Daniel Foreman-Mackey for making their codes publicly available. This research made use of Astropy, a community-developed core Python package for Astronomy \citep{Astropy13}. AHWK, MLMC and EJT acknowledge support from NASA through Hubble Fellowship grants HST-HF-51323.01-A, HST-HF-51337.01-A, and HST-HF-51316.01-A, awarded by the Space Telescope Science Institute, which is operated by the Association of Universities for Research in Astronomy, Inc., for NASA, under contract NAS 5-26555. Furthermore, MLMC acknowledges financial support from the European Research Council (ERC-StG-335936). EJT acknowledges support by a Giacconi Fellowship. KVJ was supported by National Science Foundation grant AST-1312196. SM acknowledges the new LP of the PSB.

\bibliographystyle{apj}

\end{document}